\documentclass[prd,%
preprint,%
tightenlines,showpacs,%
nofootinbib,%
superscriptaddress,%
amsfonts,amsmath]{revtex4}

\usepackage{hyperref}
\usepackage{color}

\begin{document}

\title{Cosmological self-tuning and local solutions\\
in generalized Horndeski theories}

\author{Eugeny~Babichev} \email{eugeny.babichev@th.u-psud.fr}
\affiliation{Laboratoire de Physique Th\'eorique d'Orsay,
B\^atiment 210, Universit\'e Paris-Sud 11, F-91405 Orsay Cedex,
France}
\affiliation{UPMC-CNRS, UMR7095,
Institut d'Astrophysique de Paris,
${\mathcal{G}}{\mathbb{R}}\varepsilon{\mathbb{C}}{\mathcal{O}}$,
98bis boulevard Arago, F-75014 Paris, France}

\author{Gilles~\surname{Esposito-Far\`ese}} \email{gef@iap.fr}
\affiliation{UPMC-CNRS, UMR7095,
Institut d'Astrophysique de Paris,
${\mathcal{G}}{\mathbb{R}}\varepsilon{\mathbb{C}}{\mathcal{O}}$,
98bis boulevard Arago, F-75014 Paris, France}

\begin{abstract}
We study both the cosmological self-tuning and the local
predictions (inside the Solar system) of the most general
shift-symmetric beyond Horndeski theory. We first show that the
cosmological self-tuning is generic in this class of theories: By
adjusting a mass parameter entering the action, a large bare
cosmological constant can be effectively reduced to a small
observed one. Requiring then that the metric should be close
enough to the Schwarzschild solution in the Solar system, to pass
the experimental tests of general relativity, and taking into
account the renormalization of Newton's constant, we select a
subclass of models which presents all desired properties: It is
able to screen a big vacuum energy density, while predicting an
exact Schwarzschild-de~Sitter solution around a static and
spherically symmetric source. As a by-product of our study, we
identify a general subclass of beyond Horndeski theory for which
regular self-tuning black hole solutions exist, in presence of a
time-dependent scalar field. We discuss possible future
development of the present work.
\end{abstract}

\date{September 30, 2016}


\maketitle

\section{Introduction}
\label{Sec1}
The huge discrepancy of the observed value of the cosmological
constant and its various theoretical predictions is a long
standing problem of modern physics. The value of the energy
density corresponding to the cosmological constant today, as
fitted by observations using the $\Lambda$CDM model, is of order
$10^{-46}\, \text{GeV}^4$. This value, written in units of the
Planck mass~($M_\text{Pl}$) is $\sim 10^{-122}$, which is to be
compared to the naive theoretical prediction of the vacuum energy
of order of Planck energy density. In other words, the naive
predicted value of the vacuum energy density is $10^{122}$ times
greater than the observed one. The theoretical estimate of the
value of the vacuum energy comes from the existence of a
zero-point energy of the quantized fields. The zero-point energy
density formally diverges, as it contains an integral over all
momenta of a given energy in each mode. However, the application
of a cutoff at the Planck mass gives a vacuum energy density
$\rho\sim M_\text{Pl}^4$. It has been argued, however, that one
should use a different regularization scheme, which does not
break Lorentz invariance, see the review~\cite{Martin:2012bt}.
Dimensional regularization gives in particular a different
answer, $|\rho| \sim 10^8\, \text{GeV}^4$~\cite{Koksma:2011cq}.
The problem is clearly alleviated, but the discrepancy remains
nevertheless huge, i.e., the value of the vacuum energy density
predicted in this scheme is $\sim 10^{54}$ times greater than the
observed one.

Besides the above mentioned problem of zero-point energy of
quantum fluctuations, there is yet another source of a big
cosmological constant: phase transitions in the early Universe.
In particular, the electroweak symmetry breaking, through which
the gauge bosons gain their masses, is accompanied with a change
of the vacuum value of the Higgs boson. This leads, in turn, to a
change of vacuum energy density, which is estimated to be
$|\rho_\text{EW}|\sim 10^8\, \text{GeV}^4$~\cite{Martin:2012bt}.
Similar phase transition in QCD physics leads to
$|\rho_\text{QCD}|\sim 10^{-2}\,
\text{GeV}^4$~\cite{Dolgov:1997za}. Any of these predictions
leads to too large vacuum energy.

Modifying gravity by the introduction of a scalar degree of
freedom in the gravity sector is a promising attempt to solve the
cosmological constant problem. The most general scalar-tensor
theory with equations of motion up to second order in derivatives
is known as the Horndeski theory~\cite{Horndeski}, or, in modern
formulations, the
Galileons~\cite{Nicolis:2008in,Deffayet:2009wt,Deffayet:2009mn,Deffayet:2011gz}.
The absence of higher than second derivatives in the equations of
motion guarantees the absence of any Ostrogradski ghost ---~an
extra ghost degree of freedom generically associated with higher
derivatives. The opposite is not always true, however, i.e.,
equations of motion involving higher-order derivatives do not
necessary imply the appearance of an extra degree of freedom. An
extension of the Horndeski theory has indeed been constructed,
``beyond Horndeski''
theory~\cite{Zumalacarregui:2013pma,Gleyzes:2014dya,Gleyzes:2014qga,Lin:2014jga,Deffayet:2015qwa},
which leads to third-order equations of motion, but nevertheless
with only one scalar degree of freedom.\footnote{A further
extension of the beyond Horndeski theory has also been studied
in~\cite{Langlois:2015cwa,Crisostomi:2016czh,deRham:2016wji,BenAchour:2016fzp}.
We however do not consider this ``beyond beyond Horndeski''
extension in the present paper.}

It has been shown that a subclass of the Horndeski/Galileon
theory, called ``Fab~Four'', has the property of total cancellation
of a bare cosmological
constant~\cite{Charmousis:2011bf,Charmousis:2011ea}. An extension
of the Fab~Four model, which includes the beyond Horndeski terms
holds the same property~\cite{Babichev:2015qma}. In these
scenarii the metric is flat, while the scalar field has a
non-trivial configuration. Therefore this particular model cannot
be realistic, since the observed Universe contains a small but
non-zero cosmological constant. One should thus search for a
model which would be able to {\it self-tune}, i.e., to naturally
tune the large value of a bare cosmological constant to a small
observed one.
An example of such a model, in a peculiar non-linear
extension of a subclass of Horndeski model (with an
arbitrary function of the standard kinetic term plus the
``John'' term of the Fab~Four) has been presented
in~\cite{Appleby:2012rx,Linder:2013zoa}, and further
studied in~\cite{Starobinsky:2016kua}. An approach similar
to~\cite{Charmousis:2011bf,Charmousis:2011ea} has been put
forward in~\cite{Martin-Moruno:2015bda,Martin-Moruno:2015lha}, in
order to find a subclass of the Horndeski theory which brings a
bare cosmological constant down to a smaller one fixed by the
theory itself.

It is however clear that a physically viable model should not
only demonstrate its ability for self-tuning at the cosmological
level, but it should also pass local gravity tests, in particular
solar-system tests. Any considerable deviation from general
relativity (GR) inside the Solar system would rule it out, in
spite of its nice cosmological self-tuning. For instance, as
shown in~\cite{Babichev:2012re}, a kinetic coupling between the
graviton and the scalar degree of freedom leads to the appearance
of an effective matter-scalar coupling (even in the case of a
zero bare coupling). Such a coupling is dangerous for self-tuning
models, in spite of the Vainshtein mechanism (for a review
see~\cite{Babichev:2013usa}), since it may lead to a large
backreaction of the scalar field. Indeed, the value of the time
derivative of the scalar field is expected to be naturally large,
in the self-tuning scenario. At the same time, as it has been
shown for the cubic Galileon model, the induced matter-scalar
coupling is proportional to this time derivative of the scalar
field~\cite{Babichev:2012re}. Therefore one may expect
that the backreaction of the scalar onto the geometry is large,
so that solar-system tests are not passed.

It is therefore important to identify the models which produce
self-tuning to a small observed cosmological constant, but at the
same time do not spoil solar-system tests. One such example has
been studied in~\cite{Babichev:2013cya} (see also
\cite{Cisterna:2015yla,Appleby:2015ysa}): a model containing the
``John'' term of the Fab~Four not only provides an asymptotically
de~Sitter spacetime with an effective cosmological constant,
independent from the bare vacuum constant, but it also gives a
GR-like solution near a central source.

In this paper, we systematically study cosmology and local
behavior of all shift-symmetric generalized Galileon (beyond
Horndeski) models. The action we consider, defined in
Sec.~\ref{Sec2}, contains six arbitrary functions of the standard
kinetic term of a scalar field.\footnote{Note that not all
combinations of beyond Horndeski theories are free from the
Ostrogradski ghost. One cannot mix simultaneously Horndeski and
beyond Horndeski $\mathcal{L}_4$ and $\mathcal{L}_5$ terms; see
Refs.~\cite{Langlois:2015cwa,Langlois:2015skt,Crisostomi:2016tcp}
and a short summary in Sec.~\ref{Sec2} below.} The Horndeski
theory corresponds to a particular choice of two of these
functions in terms of the other four, thus reducing the space of
the general model to four arbitrary functions. We provide in
Sec.~\ref{Sec2} a translation of our action in terms of other
notations which have been used in the literature. We also show
that the Einstein equations can be significantly simplified when
they are combined with the scalar current (whose divergence gives
the scalar-field equation).

In the first part of the paper (Sec.~\ref{Sec3}), we focus on
homogeneous cosmology of beyond Horndeski models. We derive
their general field equations, and use them to discuss several
illustrative examples of self-tuning. We show, in particular,
that an extra scale in the action (besides the Planck scale) is
necessary for the model to exhibit self-tuning. This scale
however does not need to be of order of the Hubble scale today,
although such a scenario is also allowed. Moreover, the value of
the time derivative of the scalar field can also be adjusted to
have either large or small values, depending on the theory.

In the second part (Sec.~\ref{Sec4}), we select a subset of the
beyond Horndeski theories, which provides self-tuning mechanism
in cosmology \textit{and} also restores the GR behavior for the
metric around a central source. More precisely, we exhibit a
subclass of models, depending on six functions of the standard
kinetic term, admitting an exact Schwarzschild-de~Sitter solution
around a spherical mass, and therefore \textit{a priori} able to
pass local gravity tests. Three of these functions (that we call
the ``Three Graces'') contribute actively to the self-tuning
solution, i.e., to the fact that the observed cosmological
constant $\Lambda_\text{eff}$ is much smaller than the bare one
$\Lambda_\text{bare}$ entering the action. These functions need
to be related to each other in a specific way for the model to
admit a Schwarzschild-de~Sitter solution. The three extra
functions, which also need to satisfy some relations, correspond
to ``stealth'' Lagrangians: They are allowed and do contribute to
the physics of perturbations, but they do not affect neither the
background cosmological solution nor the local spherically
symmetric metric. In Sec.~\ref{Sec4B}, we show that these exact
Schwarzschild-de~Sitter solutions also describe regular black
holes, generalizing the self-tuning black hole solutions obtained
in~\cite{Babichev:2013cya} and extended
in~\cite{Kobayashi:2014eva}.

In Sec.~\ref{Sec5}, we adopt a perturbative approach to study
which models can predict a metric close enough to the
Schwarzschild solution to pass solar-system tests, while not
giving the \textit{exact} Schwarzschild-de~Sitter solutions of
Sec.~\ref{Sec4}. We also use our perturbed field equations to
show that the observed Newton's constant $G$ is generically
renormalized with respect to the bare one entering the action of
the theory, notably in the Three Graces of Sec.~\ref{Sec4}. This
causes the cosmological constant problem \textit{not} to be
solved in most of the cases, but we show that a subclass of
models presents all desired properties: It predicts a tiny
observed cosmological constant, an exact Schwarzschild-de~Sitter
metric around a central source, and no renormalization of
Newton's constant, so that the observed vacuum energy density can
be negligible with respect to the bare one entering the action.

We finally give our conclusions in Sec.~\ref{Concl}.

\section{Generalized Horndeski theories}
\label{Sec2}

Generalized Galileon Lagrangians are most conveniently written as
contractions with two (fully-antisymmetric) Levi-Civita tensors
\cite{Deffayet:2009mn,Deffayet:2010zh}. Their main property is
then manifest, namely that their field equations in flat
spacetime depend only on second derivatives of the scalar field
$\varphi$.

In four dimensions, there are six possible such Lagrangians. We
quote below their simplest definitions, followed by their much
heavier expansions in terms of contracted covariant derivatives
and curvature tensors. We use the sign conventions
of~\cite{Misner:1974qy}, notably the mostly-plus signature, and
denote as $G_{\mu\nu} \equiv R_{\mu\nu} -\frac{1}{2} R\,
g_{\mu\nu}$ the Einstein tensor. We also denote as $\varphi_\mu
\equiv \partial_\mu\varphi$ the first derivative of the scalar
field, and similarly as $\varphi_{\mu\nu} \equiv
\nabla_\nu\nabla_\mu\varphi$ its second covariant derivative.
The six generalized Galileon Lagrangians read
\begin{subequations}
\begin{eqnarray}
L_{(2,0)} &\equiv& -\frac{1}{3!}\,\varepsilon^{\mu\nu\rho\sigma}\,
\varepsilon^\alpha_{\hphantom{\alpha}\nu\rho\sigma}\, \varphi_\mu\,
\varphi_\alpha
= (\partial_\mu\varphi)^2,
\label{eqL2Gal}\\
L_{(3,0)} &\equiv& -\frac{1}{2!}\,\varepsilon^{\mu\nu\rho\sigma}\,
\varepsilon^{\alpha\beta}_{\hphantom{\alpha\beta}\rho\sigma}\,
\varphi_\mu\, \varphi_\alpha\,
\varphi_{\nu\beta}
= (\varphi_\mu)^2\, \Box\varphi - \varphi^\mu\varphi_{\mu\nu}
\varphi^\nu,
\label{eqL3Gal}\\
L_{(4,0)} &\equiv&-\varepsilon^{\mu\nu\rho\sigma}\,
\varepsilon^{\alpha\beta\gamma}_{\hphantom{\alpha\beta\gamma}\sigma}\,
\varphi_\mu\, \varphi_\alpha\,
\varphi_{\nu\beta}\, \varphi_{\rho\gamma}
\label{eqL4Gal}\\
&=&
(\varphi_\mu)^2\left(\Box \varphi\right)^2
-2\, \varphi^{\mu} \varphi_{\mu\nu}\varphi^{\nu}\, \Box \varphi
-(\varphi_\mu)^2 (\varphi_{\nu\rho})^2
+2\, \varphi^{\mu}\varphi_{\mu\nu}
\varphi^{\nu\rho}\varphi_{\rho},\\
L_{(5,0)} &\equiv&-\varepsilon^{\mu\nu\rho\sigma}\,
\varepsilon^{\alpha\beta\gamma\delta}\, \varphi_\mu\, \varphi_\alpha\,
\varphi_{\nu\beta}\, \varphi_{\rho\gamma}\,\varphi_{\sigma\delta}
\label{eqL5Gal}\\
&=&
(\varphi_\mu)^2\left(\Box \varphi\right)^3
-3\left(\varphi^{\mu} \varphi_{\mu\nu}\varphi^{\nu}\right)
\left(\Box \varphi\right)^2
-3(\varphi_\mu)^2(\varphi_{\nu\rho})^2\, \Box \varphi
\nonumber \\
&& +6\, \varphi^{\mu}\varphi_{\mu\nu}
\varphi^{\nu\rho}\varphi_{\rho}\, \Box \varphi
+2(\varphi_\mu)^2
\varphi_{\nu}^{\hphantom{\nu}\rho}
\varphi_{\rho}^{\hphantom{\rho}\sigma}
\varphi_{\sigma}^{\hphantom{\sigma}\nu}
\nonumber \\
&& +3 \left(\varphi_{\mu\nu}\right)^2
\varphi^{\rho}\varphi_{\rho\sigma}\varphi^{\sigma}
-6\, \varphi^{\mu}\varphi_{\mu\nu}\varphi^{\nu\rho}
\varphi_{\rho\sigma}\varphi^{\sigma},\\
L_{(4,1)} &\equiv& -\varepsilon^{\mu\nu\rho\sigma}\,
\varepsilon^{\alpha\beta\gamma}_{\hphantom{\alpha\beta\gamma}\sigma}\,
\varphi_\mu\, \varphi_\alpha\, R_{\nu\rho\beta\gamma}
= - 4\, G^{\mu\nu} \varphi_\mu \varphi_\nu,
\label{eqL41Gal}\\
L_{(5,1)} &\equiv& -\varepsilon^{\mu\nu\rho\sigma}\,
\varepsilon^{\alpha\beta\gamma\delta}\, \varphi_\mu\, \varphi_\alpha\,
\varphi_{\nu\beta}\,R_{\rho\sigma\gamma\delta}
\label{eqL51Gal}\\
&=&
2(\varphi_\mu)^2 R\, \Box \varphi
-2\,\varphi^{\mu} \varphi_{\mu\nu}\varphi^{\nu}\, R
-4\, \varphi^{\mu} R_{\mu\nu}\varphi^{\nu}\, \Box \varphi
\nonumber\\
&&-4 (\varphi_\mu)^2 \varphi^{\nu\rho} R_{\nu\rho}
+8\, \varphi^{\mu}\varphi_{\mu\nu}R^{\nu\rho}\varphi_{\rho}
+4\, \varphi^{\mu} \varphi^{\nu} \varphi^{\rho\sigma}
R_{\mu\rho\nu\sigma}.
\end{eqnarray}
\label{eqGalileons}
\end{subequations}
These definitions coincide with those of
\cite{Deffayet:2009mn,Deffayet:2015qwa}
for all $L_{(n,0)}$. For those involving one Riemann
tensor, $L_{(n,1)}$, we decided to simplify them by removing a
factor $(\varphi_\lambda)^2$. We shall indeed multiply below all
these Lagrangians by arbitrary functions of
$(\varphi_\lambda)^2$, therefore this extra factor was
unnecessarily heavy in definitions (\ref{eqL41Gal}) and
(\ref{eqL51Gal}). Note that when multiplying the above
Lagrangians (\ref{eqL41Gal}) and (\ref{eqL51Gal}) by arbitrary
functions of $\varphi$, $L_{(4,1)}$ was nicknamed ``John'' in the
``Fab~Four'' model~\cite{Charmousis:2011bf,Charmousis:2011ea},
while $L_{(5,1)}$ was nicknamed ``Paul''.

Generalized Horndeski theories correspond to multiplying the
above Lagrangians by arbitrary functions of both the scalar field
$\varphi$ and its standard kinetic term $(\varphi_\lambda)^2$.
[We shall recall the difference between Horndeski and
generalized Horndeski theories below Eqs.~(\ref{eqG}).]
In the present paper, we will focus on
shift-symmetric theories, whose actions do not involve any
undifferentiated $\varphi$, and we shall thus only multiply the
above Lagrangians by functions of $(\varphi_\lambda)^2$.

In the following, we choose that $\varphi$ is dimensionless, but
introduce a mass scale $M$ so that all Lagrangians have the same
dimension. The functions will depend on the dimensionless ratio
\begin{equation}
X \equiv \frac{-(\varphi_\lambda)^2}{M^2}.
\label{eqX}
\end{equation}
Note the sign, the absence of a factor $\frac{1}{2}$, and the
$1/M^2$ factor, as the notation $X$ is used with various
definitions in the literature. Our negative sign is chosen so
that $X > 0$ in cosmological situations, where the time
derivative $\dot\varphi$ of the scalar field is dominating over
its spatial derivatives.

In addition to the mass scale $M$, which will be the only one we
use in the scalar field kinetic terms, the action we consider
also depends on two other scales: the reduced Planck mass
$M_\text{Pl} \equiv (8\pi G)^{-1/2}$ (in units such that $\hbar =
c = 1$), which multiplies the Einstein-Hilbert action, and a bare
cosmological constant $\Lambda_\text{bare}$, which may be much
larger than the observed one (see Sec.~\ref{Sec3} below for a
discussion of the effective cosmological constant
$\Lambda_\text{eff}$ which is actually observed). A simple
framework would be for instance to assume that
$\Lambda_\text{bare} = \mathcal{O}\left(M_\text{Pl}^2\right)$, so
that the model depend only on two scales, $M$ and $M_\text{Pl}$.
Let us stress that the measured Newton's constant, for
instance in Cavendish experiments, is not the bare $G\equiv
1/(8\pi M_\text{Pl}^2)$ we introduce in this action, but it acquires a
renormalized value $G_\text{eff}$. Our notation $M_\text{Pl}$ and
$G$ should thus be understood as bare parameters, whose numerical
values are not known yet. We will relate them to the observed
ones in Sec.~\ref{Sec5}, for a specific class of models which
reproduces the Schwarzschild solution in the vicinity of a
spherical body.
In order not to introduce extra hidden scales in the model, we
will assume that all functions of $X$ defined below involve
dimensionless coefficients of order $\mathcal{O}(1)$.

The class of theories we are considering is thus defined by the
full action
\begin{equation}
S = \frac{M_\text{Pl}^2}{2}
\int \sqrt{-g}\left(R - 2 \Lambda_\text{bare}\right)d^4 x
+\sum_{(n,p)} \int \sqrt{-g}\, \mathcal{L}_{(n,p)} d^4 x
+S_\text{matter}[\psi, g_{\mu\nu}],
\label{eqAction}
\end{equation}
where all matter fields different from $\varphi$ (globally
denoted as $\psi$) are assumed to be universally coupled to
$g_{\mu\nu}$ but not directly to $\varphi$, and where the
generalized Horndeski Lagrangians $\mathcal{L}_{(n,p)}$ are
related to the generalized Galileon ones (\ref{eqGalileons}) by
\begin{subequations}
\begin{eqnarray}
\mathcal{L}_{(2,0)} &=& M^2 f_2(X) L_{(2,0)} = -M^4 X f_2(X),
\label{eqL2}\\
\mathcal{L}_{(3,0)} &=& f_3(X) L_{(3,0)},
\label{eqL3}\\
\mathcal{L}_{(4,0)} &=&\frac{1}{M^2} f_4(X) L_{(4,0)},
\label{eqL4}\\
\mathcal{L}_{(5,0)} &=&\frac{1}{M^4} f_5(X) L_{(5,0)},
\label{eqL5}\\
\mathcal{L}_{(4,1)} &=& s_4(X) L_{(4,1)},
\label{eqL41}\\
\mathcal{L}_{(5,1)} &=&\frac{1}{M^2} s_5(X) L_{(5,1)}
\label{eqL51}.
\end{eqnarray}
\label{eqGeneralizedHorndeski}
\end{subequations}
Since different notation is used in the literature to define
these theories, let us give a dictionary. First of all, let us
recall that the $L_{(4,1)}$ and $L_{(5,1)}$ of
Refs.~\cite{Deffayet:2009mn,Deffayet:2015qwa} were not defined as
in Eqs.~(\ref{eqL41Gal}) and (\ref{eqL51Gal}) above, but rather
as $\mathcal{L}_{(4,1)}$ and $\mathcal{L}_{(5,1)}$,
Eqs.~(\ref{eqL41}) and (\ref{eqL51}), with $s_4 = s_5 = -X$ and
$M = 1$. Second, generalized Horndeski theories were first
defined
in~\cite{Gleyzes:2014dya,Gleyzes:2014qga,Langlois:2015cwa} with a
notation mixing the $G_n\left(\varphi_\lambda^2\right)$ functions
used for the Horndeski
theory~\cite{Horndeski,Deffayet:2011gz,Kobayashi:2011nu} and new
functions $F_n\left(\varphi_\lambda^2\right)$ multiplying the
above contractions (\ref{eqL4Gal}) and (\ref{eqL5Gal}) with two
Levi-Civita tensors:
\begin{subequations}
\begin{eqnarray}
\mathcal{L}_{(2,0)} &=& G_2\left(\varphi_\lambda^2\right),
\label{eqGeneralizedHorndeskiOld2}\\
\mathcal{L}_{(3,0)} &=& G_3\left(\varphi_\lambda^2\right)
\Box\varphi +\text{tot. div.},
\label{eqGeneralizedHorndeskiOld3}\\
\mathcal{L}_{(4,0)} + \mathcal{L}_{(4,1)} &=&
G_4\left(\varphi_\lambda^2\right) R
- 2 G_4'\left(\varphi_\lambda^2\right)
\Bigl[\left(\Box\varphi\right)^2
- \varphi_{\mu\nu}\varphi^{\mu\nu}\Bigr]\nonumber\\
&&+F_4\left(\varphi_\lambda^2\right) \varepsilon^{\mu\nu\rho\sigma}\,
\varepsilon^{\alpha\beta\gamma}_{\hphantom{\alpha\beta\gamma}\sigma}\,
\varphi_\mu\,
\varphi_\alpha\, \varphi_{\nu\beta}\, \varphi_{\rho\gamma}
+\text{tot. div.},
\label{eqGeneralizedHorndeskiOld4}\\
\mathcal{L}_{(5,0)} + \mathcal{L}_{(5,1)} &=&
G_5\left(\varphi_\lambda^2\right) G^{\mu\nu} \varphi_{\mu\nu}
+\frac{1}{3} G_5'\left(\varphi_\lambda^2\right)
\Bigl[\left(\Box \varphi\right)^3
- 3\, \Box\varphi\, \varphi_{\mu\nu}\varphi^{\mu\nu}
+ 2\, \varphi_{\mu\nu}\varphi^{\nu\rho}
\varphi_\rho^{\hphantom{\rho}\mu}\Bigr]
\nonumber\\
&&+F_5\left(\varphi_\lambda^2\right) \varepsilon^{\mu\nu\rho\sigma}\,
\varepsilon^{\alpha\beta\gamma\delta}\, \varphi_\mu\, \varphi_\alpha\,
\varphi_{\nu\beta}\, \varphi_{\rho\gamma}\,\varphi_{\sigma\delta}
+\text{tot. div.},
\label{eqGeneralizedHorndeskiOld5}
\end{eqnarray}
\label{eqGeneralizedHorndeskiOld}
\end{subequations}
where $G_4'\left(\varphi_\lambda^2\right)$ and
$G_5'\left(\varphi_\lambda^2\right)$ mean the derivatives of
these functions with respect to their argument, i.e.,
$G_n'\left(\varphi_\lambda^2\right) =
dG_n\left(\varphi_\lambda^2\right)/d\left(\varphi_\lambda^2\right
) = dG_n(-M^2 X)/d(-M^2X)$. The partial integrations given in
Appendix~\ref{AppA} below imply that these functions $G_n$ and
$F_n$ are related to our $f_n$ and $s_n$ as follows:
\begin{subequations}
\begin{eqnarray}
G_2(-M^2 X) &=& -M^4 X f_2(X),\\
G_3(-M^2 X) &=& -M^2\left[X f_3(X)
+ \frac{1}{2}\int f_3(X) dX\right],\\
G_4(-M^2 X) &=& -2 M^2 X s_4(X),
\label{eqG4}\\
F_4(-M^2 X) &=& \left[-f_4(X)+4 s_4'(X)\right]/M^2,
\label{eqF4}\\
G_5(-M^2 X) &=& 4X s_5(X) + 2\int s_5(X) dX,\\
F_5(-M^2 X) &=& \left[-f_5(X) + \frac{4}{3} s'_5(X)\right]/M^4.
\end{eqnarray}
\label{eqG}
\end{subequations}
When the functions $F_{4,5} = 0$, Horndeski~\cite{Horndeski}
showed that the field equations of these models involve at most
second derivatives. When these extra functions $F_{4,5}$ are also
present (or when our $f_{4,5}$ and $s_{4,5}$ of
Eqs.~(\ref{eqGeneralizedHorndeski}) are independent), then third
derivatives of the metric $g_{\mu\nu}$ appear in the scalar-field
equation, and third derivatives of $\varphi$ in the Einstein
equations. One could thus fear that this is associated with an
extra degree of freedom, which is generally a
ghost~\cite{Ostrogradski}, and that the models are then unstable.
References~\cite{Gleyzes:2014dya,Gleyzes:2014qga,Lin:2014jga}
underlined that this is not the case. Their initial arguments
were actually inconclusive, notably because they were working in
a specific gauge where extra degrees of freedom may be hidden
(the reason being their choice of a non-generic initial value
surface corresponding to $\varphi = \text{const.}$). However, a
full-fledged Hamiltonian analysis of the particular case of
$L_{(4,0)}$, Eq.~(\ref{eqL4Gal}), without fixing any gauge, did
show that there is indeed no extra degree of
freedom~\cite{Deffayet:2015qwa}. Under the simplifying but
reasonable hypothesis that the spin-2 sector does not hide any
subtlety,
Refs.~\cite{Langlois:2015cwa,Crisostomi:2016tcp,Crisostomi:2016czh,Langlois:2015skt}
then showed that this is also the case for \textit{most} of these
generalized Horndeski models, but curiously enough, not all of
them ---~confirming thereby that previously published arguments
were incomplete. It was shown in
\cite{Langlois:2015cwa,Crisostomi:2016tcp,Langlois:2015skt} that
any combination of $\mathcal{L}_{(4,0)}$ and
$\mathcal{L}_{(4,1)}$, Eqs.~(\ref{eqL4}) and (\ref{eqL41}), or
equivalently of the functions $F_4$ and $G_4$, is free of any
extra degree of freedom, and this results also holds for any
combination of $\mathcal{L}_{(5,0)}$ and $\mathcal{L}_{(5,1)}$,
Eqs.~(\ref{eqL5}) and (\ref{eqL51}). This is still the case when
combining any $G_4$ with any $G_5$, but with $F_4 = F_5 = 0$
(i.e., within the class of Horndeski theories~\cite{Horndeski}),
as well as when combining any $F_4$ with any $F_5$, but with $G_4
= G_5 = 0$ (i.e., when considering the Lagrangian
$\mathcal{L}_{(4,0)}+ \mathcal{L}_{(5,0)}$ without their
curvature-dependent counterparts $\mathcal{L}_{(4,1)}$ nor
$\mathcal{L}_{(5,1)}$). On the other hand, there does generically
exist an extra degree of freedom when combining arbitrary
$\mathcal{L}_{(4,0)}$, $\mathcal{L}_{(4,1)}$,
$\mathcal{L}_{(5,0)}$ and $\mathcal{L}_{(5,1)}$. In the
following, we shall study the most general case, but one should
keep in mind that all four functions $f_4$, $s_4$, $f_5$ and
$s_5$ together usually correspond to unstable models.

When performing their Hamiltonian analysis in the unitary gauge,
Refs.~\cite{Gleyzes:2014dya,Gleyzes:2014qga,Lin:2014jga}
introduced yet another notation, with functions $A_n$ and $B_n$;
see Eqs.~(25)--(29) and Appendix~A of Ref.~\cite{Gleyzes:2014qga}.
Our Appendix~A below or Eqs.~(\ref{eqG}) allow us to relate them
to our $f_n$ and $s_n$ as follows:
\begin{subequations}
\begin{eqnarray}
A_2(-M^2 X) &=& -M^4 X f_2(X),\\
A_3(-M^2 X) &=& M^3 X^{3/2} f_3(X),\\
A_4(-M^2 X) &=& -M^2 X \left[X f_4(X) + 2 s_4(X)\right],\\
B_4(-M^2 X) &=& - 2M^2 X s_4(X),\\
A_5(-M^2 X) &=& M X^{3/2}\left[X f_5(X) +2 s_5(X)\right],\\
B_5(-M^2 X) &=& -4 M X^{3/2} s_5(X).
\end{eqnarray}
\label{eqA}
\end{subequations}

The field equations are obtained by varying action
(\ref{eqAction}) with respect to the metric and the scalar field.
However, since we restrict to theories which do not depend on any
undifferentiated $\varphi$, we may write the scalar field
equation as the conservation of a current. We define the
energy-momentum tensor of the scalar field and the scalar current
respectively as
\begin{subequations}
\begin{eqnarray}
\label{eqDefT}
T^{\mu\nu}&\equiv&\frac{2}{\sqrt{-g}}\,
\frac{\delta S[\varphi]}{\delta g_{\mu\nu}},\\
J^\mu&\equiv&\frac{-1}{\sqrt{-g}}\,
\frac{\delta S[\varphi]}{\delta (\partial_\mu\varphi)},
\label{eqDefCurrent}
\end{eqnarray}
\end{subequations}
where $S[\varphi] = \sum \int\sqrt{-g}\, \mathcal{L}_{(n,p)}$ is
the $\varphi$-dependent part of action (\ref{eqAction}), and the
field equations read thus
\begin{subequations}
\begin{eqnarray}
G_{\mu\nu} + \Lambda_\text{bare}\, g_{\mu\nu}
&=& \frac{T_{\mu\nu}}{M_\text{Pl}^2},
\label{eqEinstein}\\
\nabla_\mu J^\mu &=& 0,
\label{eqCurrent}
\end{eqnarray}
\label{eqFieldEqs}
\end{subequations}
where we do not write the matter contribution to
Eq.~(\ref{eqEinstein}) because we will solve these field
equations in the exterior of material bodies. This matter
contribution will anyway be useful to fix the constant of
integration of Einstein's equations.

It should be underlined that the solutions to
Eq.~(\ref{eqCurrent}) do not correspond to $J^\mu = 0$ in
general. For instance, if $\chi$ is any solution of the free
propagation equation $\Box\chi = 0$ in a given curved background,
it is clear that $\partial^\mu\chi$ may be added to $J^\mu$
without changing Eq.~(\ref{eqCurrent}). However, in the very
symmetric situations we will consider below (homogeneous and
isotropic Universe in Sec.~\ref{Sec3}, and static and spherically
symmetric solution in Sec.~\ref{Sec4}), we will see that the
precise values of some components of $J^\mu$ may be determined,
and our solutions to (\ref{eqCurrent}) will actually correspond
to imposing $J^{\underline\mu} = 0$ for some specific index
$\underline\mu$. It happens that the Einstein equations
(\ref{eqEinstein}) can be simplified a lot by combining them with
the current as
\begin{equation}
G^{\mu\nu} + \Lambda_\text{bare}\, g^{\mu\nu}
- \frac{T^{\mu\nu}}{M_\text{Pl}^2}
+ \frac{J^\mu \varphi^\nu}{M_\text{Pl}^2}.
\label{eqCombLinEinsteinCurrent}
\end{equation}
Indeed, most of the terms involving derivatives of the functions
$f_n(X)$ and $s_n(X)$ cancel in such a combination. To prove so,
a quick and naive argument is to note that if some $f'$ or $s'$
were involved in (\ref{eqCombLinEinsteinCurrent}), then some
$f''$ or $s''$ should be generated when taking the divergence of
(\ref{eqCombLinEinsteinCurrent}). But diffeomorphism invariance
of action (\ref{eqAction}) implies that
\begin{equation}
\nabla_\mu\left(G^{\mu\nu} + \Lambda_\text{bare}\, g^{\mu\nu}
- \frac{T^{\mu\nu}}{M_\text{Pl}^2}\right)
= - \frac{\varphi^\nu \nabla_\mu J^\mu}{M_\text{Pl}^2}.
\label{eqDiffeoInv}
\end{equation}
Therefore, the divergence of (\ref{eqCombLinEinsteinCurrent}) is
just equal to $J^\mu \nabla_\mu\varphi^\nu / M_\text{Pl}^2$,
which cannot contain any second derivative $f''$ nor $s''$.
This argument is however incomplete, because $T^{\mu\nu}$ also
contains ``superpotential'' terms
\begin{equation}
4\nabla_{(\rho}\nabla_{\sigma)}\left\{
\varepsilon^{\mu\rho\alpha\gamma}
\varepsilon^{\nu\sigma\beta\delta}
\varphi_\alpha\varphi_\beta
\left[
g_{\gamma\delta} s_4(X) + \varphi_{\gamma\delta} s_5(X)/M^2
\right]
\right\},
\label{eqSuperpotential}
\end{equation}
coming from the variation of the Riemann tensor in
(\ref{eqL41Gal}) and (\ref{eqL51Gal}) with respect to the metric.
When taking their divergence, such terms reduce to a Riemann
tensor contracted with the \textit{first} derivative of the
expression within the curly brackets, because of the full
antisymmetry of the Levi-Civita tensors. Therefore, although they
contain in general first and even second derivatives of the
functions $s_4$ and $s_5$, they do not generate any $s''$ in the
divergence of~(\ref{eqCombLinEinsteinCurrent}). The conclusion is
that the above combination (\ref{eqCombLinEinsteinCurrent}) of
the Einstein equations with the scalar current cancels
\textit{most} of the first derivatives of the functions $f_n(X)$
and $s_n(X)$, but not all of them. We will use it in
Secs.~\ref{Sec3} and \ref{Sec4} below, and we will see that such
first derivatives actually do cancel in two important cases. In
the homogenous and isotropic case of Sec.~\ref{Sec3}, the reason
is that we will focus on the time-time component of the Einstein
equations, i.e., $\mu=\nu = 0$ in Eq.~(\ref{eqSuperpotential}).
But in order to create first derivatives of $s_4$ or $s_5$, at
least one of the covariant derivatives $\nabla_\rho$ or
$\nabla_\sigma$ must act on these functions, therefore at least
one among $\varphi_\alpha$ or $\varphi_\beta$ must not be
differentiated any more, and should thus correspond to the only
nonvanishing component $\varphi_0 = \dot\varphi$ in this
cosmological background. In other words, one must have $\alpha =
0$ or/and $\beta = 0$ to create a derivative of $s$ in
Eq.~(\ref{eqSuperpotential}), and there will thus be two indices
$0$ contracted with the same antisymmetric tensor $\varepsilon$,
either $\mu = \alpha = 0$ or/and $\nu=\beta=0$. This explains why
(\ref{eqSuperpotential}) will not contribute to $T^{00}$ in
Sec.~\ref{Sec3} below, and why all derivatives of $f_n(X)$ or
$s_n(X)$ will be canceled in the combination
(\ref{eqCombLinEinsteinCurrent}). In the static and spherically
symmetric case of Sec.~\ref{Sec4}, we will see that some $s_4'$
and $s_5'$ do remain, but they cancel in the particular case $X =
\text{const.}$ that we will study (and they actually cancel as
soon as one assumes $X = \text{const.}$, independently of
spherical symmetry). It is indeed clear that
(\ref{eqSuperpotential}) does not contribute to any derivative of
$s_4$ nor $s_5$ if $X = \text{const.}$ It is also easy to prove
that the only derivatives of functions $f$ or $s$ entering
$T^{\mu\nu}$ are of the form $2(\varphi^\mu \varphi^\nu / M^2)
f'(X) L_{(n,p)}$ when $X = \text{const.}$ (or with $s'$ instead
of $f'$), because they come from the variation of the metric used
in the contraction $X =-g^{\mu\nu}\varphi_\mu\varphi_\nu/M^2$. On
the other hand, the scalar current $J^\mu$ contains $2
(\varphi^\mu/M^2) f'(X) L_{(n,p)}$, and no other derivative of a
function when we assume $X = \text{const.}$ Therefore, the linear
combination (\ref{eqCombLinEinsteinCurrent}) obviously cancels
the few possible $f'$ or $s'$ which occur when $X = \text{const.}$

\section{Cosmological self-tuning}
\label{Sec3}
We consider a homogeneous and isotropic Universe whose metric
takes the Friedmann-Lema\^{\i}tre-Robertson-Walker (FLRW) form
\begin{equation}
ds^2 = -dt^2 + a^2(t)\left[
\frac{dr^2}{1-kr^2}
+ r^2\left(d\theta^2 + \sin^2\theta\, d\phi^2\right)
\right],
\label{eqFLRW}
\end{equation}
the parameter $k \in \{-1,0,1\}$ determining whether the spatial
hypersurfaces are open, flat or closed, and we assume
consistently that the scalar field $\varphi$ depends only on $t$.
Then its current equation (\ref{eqCurrent}) simply reads
$\nabla_0 J^0 = 0 \Rightarrow \partial_t(a^3 J^0) = 0$, and its
solution is thus $J^0 = C_0/a^3$, where $C_0$ is a constant. This
integration constant may be neglected at late enough times, when
the scale factor $a$ becomes very large, and we will thus just
solve for $J^0 = 0$ in the following (keeping in mind that an
extra $C_0/a^3$ may be added to it).

Once the matter field equations are taken into account, i.e.,
$\nabla_\mu J^\mu = 0$ in the present case, it is well known that
only the time-time-component of the Einstein equations
(\ref{eqEinstein}) needs to be solved. Indeed, the covariant
conservation of the Einstein tensor $\nabla_\mu G^{\mu\nu} = 0$
implies $G^{ij} = - g^{ij}\left[G^{00} + \dot
G^{00}/(3H)\right]$, where a dot denotes time differentiation
and $H\equiv \dot a/a$,
therefore the spatial components of the Einstein tensor are
automatically solved once $G^{00}$ is, while all off-diagonal
components vanish identically. When taking into account
the energy-momentum of the scalar field, this relation remains
valid for $G^{\mu\nu} -T^{\mu\nu}/M_\text{Pl}^2$ instead of
$G^{\mu\nu}$, up to terms proportional to the scalar equation
$\nabla_\mu J^\mu = 0$, because of Eq.~(\ref{eqDiffeoInv})
implied by the diffeomorphism invariance of action
(\ref{eqAction}).

We thus only write below the $00$-component of the Einstein
equations and the scalar current. Actually, instead of the
Einstein equation, we choose to write the $00$-component of the
linear combination (\ref{eqCombLinEinsteinCurrent}), which
simplifies significantly its expression, and we then display
$J^0/(2 M^2 \dot\varphi)$:
\begin{subequations}
\begin{eqnarray}
3\left(H^2+\frac{k}{a^2}\right)-\frac{M^4}{M_\text{Pl}^2}X f_2
+ 6H^2 \frac{M^2}{M_\text{Pl}^2} X^2 f_4&&\nonumber\\
+ 12\left(H^2-\frac{k}{a^2}\right) \frac{M^2}{M_\text{Pl}^2} X s_4
-12 \frac{H^3 M}{M_\text{Pl}^2}\left[X^{5/2} f_5+ 2 X^{3/2} s_5\right]
&=& \Lambda_\text{bare},
\label{eqCosmo00k}\\
\left[X f_2\right]' - 3\frac{H}{M}\left[X^{3/2} f_3\right]'
+ 6\left(\frac{H}{M}\right)^2\left[X^2 f_4\right]'
- 6\left(\frac{H}{M}\right)^3\left[X^{5/2} f_5\right]'\nonumber\\
+ 12\, \frac{(H^2 + k/a^2)}{M^2}\, \left[X s_4\right]'
- 12\,\frac{H(H^2+ k/a^2)}{M^3}\,\left[X^{3/2} s_5\right]'
&=& 0,
\label{eqJ0k}
\end{eqnarray}
\label{eqCosmok}
\end{subequations}
where a prime denotes differentiation with respect to $X$. [If
the integration constant of $J^0 = C_0/a^3$ is not neglected,
then this adds $C_0 \dot\varphi/(a^3 M_\text{Pl}^2)$ to the
right-hand side (r.h.s.) of Eq.~(\ref{eqCosmo00k}) and $C_0/(2
a^3 M^2 \dot\varphi)$ to the r.h.s. of Eq.~(\ref{eqJ0k}).] Note
the different signs of the spatial curvature contribution $k/a^2$
in the various terms. This comes from the fact that it does not
enter with the same weight in the Einstein equation and the
scalar current. For instance, this curvature contribution happens
to vanish for the $s_5$ term in the linear combination
(\ref{eqCosmo00k}), whereas Eq.~(\ref{eqJ0k}) shows that it was
initially present both in $T^{00}$ and $J^0$.

Note also that we do \textit{not} assume $\dot H= 0$ nor
$\ddot\varphi = 0$ in these equations: They are fully general in
FLRW. The third usual cosmological equation, involving $\dot H$,
is a consequence of the above two. However, the solutions we will
exhibit below will actually correspond to a de~Sitter Universe
with $3 H^2= \Lambda_\text{eff} = \text{const.}$ and $\dot\varphi
= \text{const.}$

We obtained Eqs.~(\ref{eqCosmok}) by two independent methods:
First by deriving the full covariant equations and specifying
them to metric (\ref{eqFLRW}); and second the ``minisuperspace''
technique, in which the form (\ref{eqFLRW}) is imposed directly
within the action (with an arbitrary $g_{00}(t) dt^2$ instead of
$-dt^2$), and varying it with respect to the two fields
$g_{00}(t)$ and $\dot\varphi(t)$. [The non-independent
$rr$-component of the Einstein equations is also immediate to
obtain by varying this action with respect to $a(t)$.]

In the following, we shall assume that the spatial curvature
vanishes, $k = 0$. In such a case, Eqs.~(\ref{eqCosmok})
simplify even more, and our two cosmological equations take
elegant similar forms:
\begin{subequations}
\begin{eqnarray}
-X f_2
+ 6\left(\frac{H}{M}\right)^2\left[X^2 f_4
+ 2 X s_4\right]\hphantom{'}\nonumber\\
-12 \left(\frac{H}{M}\right)^3
\left[X^{5/2} f_5+ 2 X^{3/2} s_5\right]\hphantom{'}
&=& \frac{M_\text{Pl}^2}{M^4}\left(\Lambda_\text{bare} - 3 H^2\right),
\label{eqCosmo00}\\
\left[X f_2\right]' - 3\frac{H}{M}\left[X^{3/2} f_3\right]'
+ 6\left(\frac{H}{M}\right)^2\left[X^2 f_4+ 2 X s_4\right]'\nonumber\\
- 6\left(\frac{H}{M}\right)^3
\left[X^{5/2} f_5+ 2 X^{3/2} s_5\right]' &=& 0.
\label{eqJ0}
\end{eqnarray}
\label{eqCosmo}
\end{subequations}
Note that the combinations of functions entering the various
square brackets are precisely those which correspond to the $A_n$
notation of Eqs.~(\ref{eqA}). This comes from the fact that this
notation was introduced while performing an ADM decomposition in
the unitary gauge, which is close to the above factorization of
powers of $H/M$, in this cosmological context where all fields
only depend on time.

This writing immediately exhibits some singular limiting cases,
in the present cosmological framework. First, any linear
combination of $f_2\propto 1/X$, $f_3\propto X^{-3/2}$,
$f_4\propto X^{-2}$, $f_5\propto X^{-5/2}$, $s_4\propto 1/X$, and
$s_5\propto X^{-3/2}$, obviously gives a trivially satisfied
$0=0$ equation for the current (\ref{eqJ0}), while the Einstein
equation (\ref{eqCosmo00}) becomes fully independent from $X$.
Therefore, such models fail at predicting which cosmological
value $X$ should take (let us call it $X_c$), and $\dot\varphi_c
= M \sqrt{X_c}$ is thus free. On the other hand, such models
anyway do predict a specific value of $H$ (when the signs of the
various terms are consistent), and thereby of the observed
cosmological constant $\Lambda_\text{eff} = 3 H^2$. It should
however be noted that two Lagrangians among those are
particularly trivial. The first one is $\mathcal{L}_{(2,0)}$ when
$f_2\propto 1/X$. Then Eq.~(\ref{eqL2}) shows that this
Lagrangian is nothing else than a second bare cosmological
constant ---~which can obviously almost compensate
$\Lambda_\text{bare}$, but this would be equivalent to assuming
that there is no large bare cosmological constant in our initial
action (\ref{eqAction}). The second trivial case is when
$f_3\propto X^{-3/2}$ in the Lagrangian $\mathcal{L}_{(3,0)}$,
Eq.~(\ref{eqL3}). Then it happens to be a total derivative, and
thereby not to contribute to any field equation, even in generic
non-symmetric situations. We have indeed
$-\nabla_\mu\left[\varphi^\mu/\sqrt{-\varphi_\lambda^2}\right]=
\left(-\varphi_\lambda^2\right)^{-3/2}
\left[\varphi_\mu^2\, \Box\varphi
- \varphi^\mu\varphi_{\mu\nu} \varphi^\nu\right]
= \mathcal{L}_{(3,0)}/M^3$.
Aside from these two trivial cases, the other $f_4\propto
X^{-2}$, $f_5\propto X^{-5/2}$, $s_4\propto 1/X$ and $s_5\propto
X^{-3/2}$ are perfectly allowed, notably when they are combined
with other terms which are not of this specific form. One should
just keep in mind that these limiting cases do not contribute to
the current (\ref{eqJ0}), and this constrains the form of the
other terms added to them. For instance, if one considers the sum
of $\mathcal{L}_{(2,0)}$ with one of these limiting cases, then
Eq.~(\ref{eqJ0}) implies that one must have $\left(X f_2\right)'
= 0$, i.e., that $\mathcal{L}_{(2,0)}$ behaves a second
cosmological constant at least around the background value $X =
X_c$, and this could not be called an actual ``self-tuning''.

Another class of non-fully predictive models is also exhibited by
Eqs.~(\ref{eqCosmo}). When $f_3$, $f_4$, $f_5$, $s_4$ and $s_5$
are monomials (some of them possibly vanishing) such that
\begin{equation}
(X f_3)^6 \propto (X f_4 + 2 s_4)^3 \propto (X f_5 + 2 s_5)^2,
\label{eqProp}
\end{equation}
and $f_2 = 0$, or when
\begin{equation}
X^2 f_4 + 2X s_4 = \text{const.} \quad\text{or/and}\quad
X^{5/2} f_5 + 2 X^{3/2} s_5 = \text{const.},
\label{eqFconst}
\end{equation}
and $f_2 = f_3 = 0$ (or the trivial $f_2\propto 1/X$ or
$f_3\propto X^{-3/2}$ mentioned above), then the two equations
depend on only one variable, which is the product of $H$ by a
given function of~$X$. This variable is $H X^{3/2} f_3$ in the
case of Eq.~(\ref{eqProp}) [or $H(X^2 f_4 + 2 X s_4)^{1/2}
\propto H (X^{5/2} f_5 + 2 X^{3/2} s_5)^{1/3}$ if $f_3 = 0$], and
simply $H$ in the case of Eq.~(\ref{eqFconst}). Therefore, $H$
and $X$ cannot be both predicted in these particular models.
Although this \textit{a priori} means that some kinetic term
vanishes, this does not necessarily rule out such models. Indeed,
their Cauchy problem may be well posed around slightly different
backgrounds, and their non-predictivity may thus happen only when
assuming an exact FLRW metric (\ref{eqFLRW}). It remains that we
must disregard them in the present FLRW framework, since they let
undetermined a combination of $H$ and $X = (\dot\varphi/M)^2$.
Note that in the case of Eq.~(\ref{eqFconst}), the current
equation (\ref{eqJ0}) is trivially satisfied, but not in the case
of Eq.~(\ref{eqProp}), for which (\ref{eqCosmo}) become now two
independent equations of a single variable. To make them
consistent with each other, it is thus necessary to impose a very
specific value of $M$, depending on the other parameters of the
model.

Coming back now to the \textit{generic} case of
Eqs.~(\ref{eqCosmo}), we find that self-tuning is possible with
\textit{any} combination of at least two Lagrangians
$\mathcal{L}_{(n,p)}$, Eqs.~(\ref{eqGeneralizedHorndeski}).
Indeed, although the physical meaning of these equations is to
predict $H$ and $X$ in a given theory [defined by fixed values of
$M$, $M_\text{Pl}$, $\Lambda_\text{bare}$ and fixed functions
$f_n(X)$ and $s_n(X)$], we may also consider them as equations
determining $M$ and $X$ in terms of the \textit{observed} value
of $H$ and the other parameters and functions defining the model.
Therefore, it suffices to tune the mass scale $M$ to an
appropriate numerical value to get the observed $3 H^2 =
\Lambda_\text{eff}$ as a solution, in spite
of the large bare cosmological constant which is introduced in
action (\ref{eqAction}). Of course, the fact that we need to tune
$M$ means that we are actually introducing by hand some
information about $H$ in our action. However, this parameter $M$
defines the dynamics of the scalar field $\varphi$, and has thus
no relation with the quantum vacuum energy density. It is
therefore less problematic to tune its value. Moreover, we will
see that $M$ does not need to be of the same order of magnitude
as $H$: There exist self-tuning models with all kinds of values
for $M$, e.g., $M \ll H$, $M \approx H$, $M \gg H$, and even
trans-Planckian $M \gg M_\text{Pl}$ [but note that this scale $M$
is not the energy of any localized wave packet, nor even a mass,
but simply a dimensionful scale necessary to define the
Lagrangians (\ref{eqGeneralizedHorndeski})]. It is notably
possible to have $M$ of an intermediate magnitude, larger than
the heaviest masses of the Standard Model of particle physics but
smaller than the Planck mass.

The only order of magnitude that $M$ cannot consistently take is
$M \approx \left(M_\text{Pl}^2 \Lambda_\text{bare}\right)^{1/4}$.
Indeed, if this were the case, then Eqs.~(\ref{eqCosmo}) would
involve only this scale and dimensionless numbers assumed to be
of order 1, therefore the predicted $H$ would also be generically
of this order, i.e., much larger than the observed one. Actually,
if one enforces $M = \mathcal{O}\left(M_\text{Pl}^2
\Lambda_\text{bare}\right)^{1/4}$ in Eqs.~(\ref{eqCosmo}), one
can find models which would still be consistent with a small $H$,
but they are in the ``non-fully predictive'' class mentioned in
Eq.~(\ref{eqProp}) above. They indeed predict the value of a
product of $H$ with a function of $X$, but nothing else.
Therefore, if it happens that this function of $X$ takes a large
value in our Universe, then this will correspond to a small
observed $H$, but any other value of $H$ would have also been
possible. In such non-predictive models, the absence of a second
scale in the action is thus compensated by the random scale that
can take the function of $X$. Let us just quote one example in this
non-fully predictive class. If $f_3 = 1$, $f_4 = X$ and all other
functions vanish, then both field equations (\ref{eqCosmo})
depend only on the product $H X^{3/2}$, and they are consistent
with each other only if one imposes $M^4 = \frac{8}{3}\,
M_\text{Pl}^2 \Lambda_\text{bare} =
\mathcal{O}\left(M_\text{Pl}^2 \Lambda_\text{bare}\right)$. Then
they predict $H X^{3/2} = \frac{1}{4}\, M$, so that $H$ may be
small if $X$ happens to be large, but a large value of $H$ is
equally allowed by the same equations. Although amusing, we shall
disregard such single-scale non-fully predictive models in the
following. Those we will focus on will therefore necessarily
involve a second scale $M$, either large or small with respect to
the vacuum energy scale $\left(M_\text{Pl}^2
\Lambda_\text{bare}\right)^{1/4}$.

As mentioned above, self-tuning is possible when at least two
Lagrangians $\mathcal{L}_{(n,p)}$ are present in the theory, and
we did study systematically all possible combinations of them.
Let us just quote here some examples, to illustrate the diversity
of their predictions. We shall see in Secs.~\ref{Sec4} and
\ref{Sec5b} that the subclass of models $\mathcal{L}_{(2,0)}+
\mathcal{L}_{(4,0)} +\mathcal{L}_{(4,1)}$, that we may call the
``Three Graces'', is the most interesting. Let us therefore focus
on this subclass for the present illustration, and to simplify,
let us choose monomials $f_2 = k_2 X^\alpha$, $f_4 = k_4 X^\beta$
and $s_4 = \kappa_4 X^\gamma$, where $k_2$, $k_4$ and $\kappa_4$
are $\mathcal{O}(1)$ dimensionless constants, whose signs are
imposed by the two cosmological equations (\ref{eqCosmo}). As
underlined above, we must choose $\alpha \neq -1$ otherwise the
$\mathcal{L}_{(2,0)}$ Lagrangian, Eq.~(\ref{eqL2}), is another
bare cosmological constant. We see that the $s_4$ function
behaves exactly as $f_4$ if $\gamma = \beta+1$ and $\kappa_4 =
k_4/2$. If both $f_4$ and $s_4$ are assumed to contribute with
the same order of magnitude in these equations,\footnote{The
Horndeski combination corresponds to $F_4 = 0$ in
Eq.~(\ref{eqF4}), i.e., to $f_4 = 4 s_4'$, therefore to $f_4 = k_4
X^{\beta}$ and $s_4 = \frac{1}{4} k_4 X^{\beta+1}/(\beta+1)$ in
the present monomial case. For the cosmological background, it
behaves thus as if there were only $f_2$ and $f_4$ with $k_4$
replaced by $(2\beta+3)k_4/(2\beta+2)$.} then we \textit{need}
$\gamma = \beta+1$, and everything behaves as if there were only
$f_2$ and $f_4$ with $k_4$ replaced by $k_4+2\kappa_4$. We may
thus consider only the case $\mathcal{L}_{(2,0)}+
\mathcal{L}_{(4,0)}$. Then Eqs.~(\ref{eqCosmo}) imply
\begin{subequations}
\begin{eqnarray}
M &\propto& \left[H^{-(\alpha+1)}
\left(M_\text{Pl}^2 \Lambda_\text{bare}
\right)^{(\alpha-\beta-1)/2}\right]^{1/(\alpha-2\beta-3)},
\label{eqM}\\
X &\propto& \left[H^4/
\left(M_\text{Pl}^2 \Lambda_\text{bare}
\right)\right]^{1/(\alpha-2\beta-3)},\\
\dot\varphi &\propto& \left[H^{1-\alpha}
\left(M_\text{Pl}^2 \Lambda_\text{bare}
\right)^{(\alpha-\beta-2)/2}\right]^{1/(\alpha-2\beta-3)},
\label{eqOrderOfMagnitudePhiDot}
\end{eqnarray}
\label{eqOrderOfMagnitude}
\end{subequations}
with $\mathcal{O}(1)$ numerical factors depending on the
constants $k_2$ and $k_4$ and the exponents $\alpha$ and $\beta$.
In the realistic case where $H \ll \left(M_\text{Pl}^2
\Lambda_\text{bare}\right)^{1/4}$, we deduce thus that $M$, $X$
and $\dot\varphi$ may be independently small or large depending
on the positive or negative signs of the exponents of $H$ in
Eqs.~(\ref{eqOrderOfMagnitude}). We find that $M$ is small (with
respect to $M_\text{Pl}$) if $\alpha < -1$ and $\beta <
(\alpha-3)/2$, or if $\alpha > -1$ and $\beta > (\alpha-3)/2$. On
the other hand, $X$ is small (with respect to 1) if $\beta <
(\alpha-3)/2$. Finally, $|\dot\varphi|$ is small (with respect to
$M_\text{Pl}$) if $\alpha < 1$ and $\beta < (\alpha-3)/2$, or if
$\alpha > 1$ and $\beta > (\alpha-3)/2$. We quote below some even
more specific examples to illustrate that these quantities can be
independently large or small. Note that $M$ must be either large
or small, but never of the order of magnitude of
$\left(M_\text{Pl}^2 \Lambda_\text{bare}\right)^{1/4}$, because
$\alpha = -1$ is forbidden. As underlined above, this behavior is
actually valid for all combinations of Lagrangians
(\ref{eqGeneralizedHorndeski}) [unless we are in a limiting case
which cannot predict the value of $X$]. Similarly, $X$ may be
either large or small, but never order 1 in the present
$\mathcal{L}_{(2,0)}+ \mathcal{L}_{(4,0)}$ model, otherwise this
would correspond to some infinite exponent $\alpha$ or $\beta$.
However, some other combinations of Lagrangians do allow for $X =
\mathcal{O}(1)$, notably the
$\mathcal{L}_{(4,0)}+\mathcal{L}_{(4,1)}$ and
$\mathcal{L}_{(5,0)}+\mathcal{L}_{(5,1)}$ cases (without any
$\mathcal{L}_{(2,0)}$), where $X = \mathcal{O}(1)$ is actually
implied by the field equations. [These
$\mathcal{L}_{(4,0)}+\mathcal{L}_{(4,1)}$
and $\mathcal{L}_{(5,0)}+\mathcal{L}_{(5,1)}$ combinations are
also the only ones for which it is impossible to choose a small
mass scale $M$, as one finds $M \propto H^{-1}$ in the first case
and $M \propto H^{-3}$ in the second.] Finally, note that
$|\dot\varphi| \sim \left(M_\text{Pl}^2
\Lambda_\text{bare}\right)^{1/4}$ is possible in
Eq.~(\ref{eqOrderOfMagnitudePhiDot}), if one chooses
$\alpha = 1$, i.e., $f_2(X) = k_2 X$.

Let us quote some more specific examples to illustrate their
$\mathcal{O}(1)$ numerical factors and the relative sizes of
their predictions.

For $f_2 = -1$ and $f_4 = 1$, we get (while neglecting $3H^2 =
\Lambda_\text{eff}$ with respect to $\Lambda_\text{bare}$)
\begin{subequations}
\begin{eqnarray}
M &=& \left(8 H^2 M_\text{Pl}^2 \Lambda_\text{bare}\right)^{1/6}
\quad \Leftrightarrow\quad
\Lambda_\text{eff} = \frac{3 M^6}{8 M_\text{Pl}^2
\Lambda_\text{bare}},\\
X &=& \frac{1}{6}\left(\frac{M_\text{Pl}^2
\Lambda_\text{bare}}{H^4}\right)^{1/3},\\
|\dot\varphi| &=& \frac{1}{\sqrt{6}}
\left(\frac{M_\text{Pl}^2 \Lambda_\text{bare}}{H}\right)^{1/3}.
\end{eqnarray}
\label{eqExample1}
\end{subequations}
Therefore, $M$ is small but both $X$ and $|\dot\varphi|$ are
large in this model.

For $f_2 = -X^2$ and $f_4 = 1$, we get
\begin{subequations}
\begin{eqnarray}
M &=& \frac{4 \sqrt{10}\, H^3}{M_\text{Pl} \sqrt{\Lambda_\text{bare}}}
\quad \Leftrightarrow\quad
\Lambda_\text{eff} = \frac{3}{2}\left(\frac{M^2 M_\text{Pl}^2
\Lambda_\text{bare}}{20}\right)^{1/3},\\
X &=& \frac{M_\text{Pl}^2 \Lambda_\text{bare}}{40 H^4},\\
|\dot\varphi| &=& 2 H.
\end{eqnarray}
\label{eqExample2}
\end{subequations}
Therefore, $X$ is large but both $M$ and $|\dot\varphi|$ are
small in this model.

We shall see in Sec.~\ref{Sec5b} that the models with $f_2 = k_2
X^\alpha$ and $f_4 = k_4 X^{-5/2}$ (or/and $s_4 = \kappa_4
X^{-3/2}$) have the quite interesting property that the observed
Newton's constant $G$ is not renormalized ---~i.e., that it is
equal to the bare one entering action (\ref{eqAction}). Let us
just quote here one example among them, say the nicely symmetric
$f_2 = -X^{-3/2}$ and $s_4 = X^{-3/2}$. We get
\begin{subequations}
\begin{eqnarray}
M &=& 2\sqrt{3}\, H
\quad \Leftrightarrow\quad
\Lambda_\text{eff} = M^2/4,\\
X &=& \left(\frac{24\, H^4}{M_\text{Pl}^2
\Lambda_\text{bare}}\right)^2,\\
|\dot\varphi| &=& \frac{48\sqrt{3}\,
H^5}{M_\text{Pl}^2 \Lambda_\text{bare}}.
\end{eqnarray}
\label{eqExample3}
\end{subequations}
Therefore, $M$, $X$ and $|\dot\varphi|$ are all small in this
model. Note that obtaining all $M$, $X$ and $|\dot\varphi|$ small
and keeping an unrenormalized Newton's constant are fully
independent properties. For instance, the model $f_2 = 1$ and
$f_4 = X^{-5/2}$ does also predict an unrenormalized $G$ but
gives $M \propto H^{-1/2}$ large while $X \propto H^2$ and
$|\dot\varphi| \propto H^{1/2}$ are small. Note finally that
negative powers are not problematic in our present cosmological
context, since the background value of $\dot\varphi^2 = M^2 X$
does not vanish, and that $X$ should thus not pass through zero.
Negative powers are obviously a much more serious issue when
considering Horndeski theories around a vanishing scalar
background, and some surprising results of the
literature~\cite{DeFelice:2015sya} are actually related to such
negative powers. Perturbations of the scalar field are very
probably ill-defined in such cases.

The above examples show that the magnitude of $|\dot\varphi|$
depends crucially on the considered model. However, let us
underline that the energy-momentum tensor $T^{\mu\nu}$ of this
scalar field is \textit{always} of order $M_\text{Pl}^2
\Lambda_\text{bare}\, g^{\mu\nu}$. Indeed, self-tuning means
\textit{by construction} that this energy-momentum tensor must
almost compensate the bare cosmological constant in the Einstein
equations (\ref{eqCombLinEinsteinCurrent}). Therefore, whatever
the values of $M$ or $\dot\varphi$, we have anyway large
$\mathcal{O}(M_\text{Pl})$ scalar effects, at least in the
background.

We end this Section by mentioning another class of particular
cases: models predicting an observed $3H^2 = \Lambda_\text{eff}$
which is fully independent from $\Lambda_\text{bare}$. In the
$\mathcal{L}_{(2,0)} +\mathcal{L}_{(4,0)}$ subclass of models,
Eq.~(\ref{eqM}) shows that this happens when $\alpha = \beta +1$,
i.e., when $f_2 = k_2 X^\alpha$ and $f_4 = k_4 X^{\alpha-1}$.
Indeed, one then gets $\Lambda_\text{eff}\approx M^2$, up to an
$\mathcal{O}(1)$ numerical coefficient depending on the
dimensionless constants $k_2$, $k_4$ and $\alpha$. Therefore, for
a given theory with fixed $M$, the observed cosmological constant
$\Lambda_\text{eff}$ will remain unchanged even after a phase
transition modifying $\Lambda_\text{bare}$. Note however that
this necessarily means that the observed Hubble scale $H$ is
actually introduced by hand in the action, via the mass scale $M
\approx H$, since $H$ is independent of $\Lambda_\text{bare}$.
Some fine-tuning is thus still required in this subclass of
models, although it is now for a mass scale entering the
generalized Horndeski Lagrangians, and no longer for a vacuum
energy whose quantum prediction cannot have the observed order of
magnitude.

As underlined above Eqs.~(\ref{eqOrderOfMagnitude}),
$\mathcal{L}_{(4,0)}$ and $\mathcal{L}_{(4,1)}$ give almost
identical field equations. Predicting an observed
$\Lambda_\text{eff}$ independent from $\Lambda_\text{bare}$ is
thus obviously possible too in the $\mathcal{L}_{(2,0)}
+\mathcal{L}_{(4,1)}$ subclass of models with $\alpha = \gamma$,
i.e., with $f_2 = k_2 X^\alpha$ and $s_4 = \kappa_4 X^\alpha$.
The particular case $\alpha = 0$, corresponding to $f_2 = -1$ and
$s_4 = 1$, was actually the first model found with this property,
in Ref.~\cite{Babichev:2013cya}. Another example of this kind is
given in Eqs.~(\ref{eqExample3}) above, where $\Lambda_\text{eff}
= M^2/4$ is indeed independent from $\Lambda_\text{bare}$. This
behavior can also be obtained in most other combinations of
Lagrangians $\mathcal{L}_{(n,p)}$, for instance with $f_2 = k_2
X^\alpha$ and either $f_3 = k_3 X^{\alpha - 1/2}$, or $f_5 = k_5
X^{\alpha - 3/2}$, or $s_5 = \kappa_5 X^{\alpha - 1/2}$ (all
other functions being assumed to vanish). The only particular
cases for which it is not possible to predict
$\Lambda_\text{eff}$ independent from $\Lambda_\text{bare}$ are
again the $\mathcal{L}_{(4,0)}+\mathcal{L}_{(4,1)}$
and$\mathcal{L}_{(5,0)}+\mathcal{L}_{(5,1)}$ combinations, that
we already mentioned in the paragraph below
Eqs.~(\ref{eqOrderOfMagnitude}): One always predicts
$\Lambda_\text{eff} \propto \Lambda_\text{bare}$ in the first
case, while $\Lambda_\text{eff} \propto
\Lambda_\text{bare}^{2/3}$ in the second.

\section{Schwarzschild-de~Sitter solutions}
\label{Sec4}

\subsection{Self-tuning solutions around a spherical body}
\label{Sec4A}
We now consider the same models as above, with the same
cosmological behavior at large distances, but we study their
predictions in the vicinity of a spherical and static massive
body. Are they consistent with the Schwarzschild metric, which is
very well tested at the first post-Newtonian order in the solar
system? We do know that these models generically exhibit a
Vainshtein mechanism, which reduces the observable scalar-field
effects at small enough distances. But in the present self-tuning
context, we saw in Sec.~\ref{Sec3} that some quantities (like
$\dot\varphi$) can take extremely large values, therefore the
backreaction of the scalar field can \textit{a priori} fully
change the behavior of the metric, and solar-system tests are not
guaranteed to be passed. Actually, one might even fear that none
of these models is consistent with local tests, in spite of the
Vainshtein mechanism.

A large number of works has been devoted in the literature to the
Vainshtein mechanism in Galileon theories. However, most of the
studies assumed a time-independent scalar field, see
e.g.~\cite{Babichev:2009ee,Kimura:2011dc,Narikawa:2013pjr,
DeFelice:2011th,Kase:2013uja,Koyama:2013paa,Charmousis:2015aya}
(for a recent review on the Vainshtein mechanism
see~\cite{Babichev:2013usa}). The Vainshtein mechanism
with a time-dependent scalar has been considered
in~\cite{Babichev:2011iz,Babichev:2012re}, while
Ref.~\cite{Kobayashi:2014ida} studied it in a subclass
of beyond Horndeski theories (with $\mathcal{L}_{(5,0)}=
\mathcal{L}_{(5,1)} =\Lambda_\text{bare} = 0$).

Our approach is quite different in the present Section. We shall
scan the whole class of generalized Horndeski theories to look
for a subclass which (i)~is able to screen a huge cosmological
constant $\Lambda_\text{bare}$, and (ii)~reproduces the
\textit{exact} Schwarzschild-de~Sitter solution of GR with a
small but non-vanishing observed cosmological constant
$\Lambda_\text{eff}$.

We choose to work in Schwarzschild coordinates
\begin{equation}
ds^2 = -e^{\nu(r)} dt^2 + e^{\lambda(r)} dr^2
+ r^2 \left(d\theta^2 + \sin^2\theta\, d\phi^2\right),
\label{eqSchwarzschildCoords}
\end{equation}
and we consider a scalar field of the form
\begin{equation}
\varphi = \dot\varphi_c t + \psi(r),
\label{scalaransatz}
\end{equation}
where $\dot\varphi_c$ is now assumed to be a constant [contrary
to Eqs.~(\ref{eqCosmok}) and (\ref{eqCosmo}) above, which were
valid for any time dependence]. This ansatz (\ref{scalaransatz}),
which may be considered as the first approximation of a more
complicated time evolution, allows us to separate time and radial
variables in all field equations because of the shift-symmetry of
the theory. Indeed, since an undifferentiated $\varphi$ cannot
appear in these field equations, time derivatives are transformed
into the constant $\dot\varphi_c$ (or $0$), and we thus get only
\textit{ordinary} differential equations with respect to the
radial coordinate. We also define the dimensionless ratio
\begin{equation}
q \equiv \frac{\dot\varphi_c}{M},
\end{equation}
and the cosmological value of the standard kinetic term
(\ref{eqX}) reads thus $X_c = q^2 = \text{const.}$

In the previous section, we saw that the $00$-component of the
Einstein equations and the scalar equation were enough to solve
all field equations in FLRW. In the present static and
spherically symmetric situation, three equation become necessary
and sufficient. Indeed, the covariant conservation of the
Einstein tensor $\nabla_\mu G^{\mu\nu} = 0$ implies now $2
e^\lambda G_{\theta\theta}/r^3 =\partial_r G_{rr} +(2/r -
\lambda' +\nu'/2)G_{rr} + \nu' e^{\lambda-\nu} G_{00}/2$, where a
prime denotes radial differentiation, therefore the angle-angle
components $G_{\phi\phi} = \sin^2\theta\, G_{\theta\theta}$ are
automatically solved once the $00$ and $rr$ components are, while
all off-diagonal components vanish identically. As before, when
taking into account the energy-momentum of the scalar field, this
remains valid for $G^{\mu\nu} -T^{\mu\nu}/M_\text{Pl}^2$ instead
of $G^{\mu\nu}$, up to terms proportional to the scalar equation
$\nabla_\mu J^\mu = 0$, because of Eq.~(\ref{eqDiffeoInv})
implied by the diffeomorphism invariance of action
(\ref{eqAction}).

We give in Appendix~\ref{AppB} the two relevant Einstein
equations and the scalar current. Actually, we also simplified
the $rr$-Einstein equation by combining it with the scalar
current as in Eq.~(\ref{eqCombLinEinsteinCurrent}). In the present
spherically symmetric case with $\dot\varphi = \text{const.}$,
the scalar equation $\nabla_\mu J^\mu = 0$ simply reads
$\partial_r (r^2J^r) = 0$. Its solution is thus in general $J^r =
C_r/r^2$, where $C_r$ is an integration constant. However,
since we assume that there is no bare matter-scalar coupling in
action (\ref{eqAction}), the scalar field does not have any
source term even within matter, therefore this integration
constant must vanish (otherwise the scalar field would be
singular at $r = 0$).

Since Eqs.~(\ref{B1})--(\ref{B3}) are quite heavy, we checked
them again by two independent methods: First by deriving
the full covariant equations and specifying them to metric
(\ref{eqSchwarzschildCoords}); and second the ``minisuperspace''
technique, in which the form (\ref{eqSchwarzschildCoords}) is
imposed directly within the action, and then one varies it
with respect to the three fields $\nu(r)$, $\lambda(r)$ and
$\varphi'(r) = \psi'(r)$.

Our aim is now to exhibit a subclass of models which is
consistent with an exact Schwarzschild-de~Sitter metric.
We therefore impose so in Eqs.~(\ref{B1})--(\ref{B3}),
by enforcing the metric to take the form
(\ref{eqSchwarzschildCoords}) with
\begin{equation}
e^\nu = e^{-\lambda} = 1-\frac{r_s}{r} - (H r)^2,
\label{eqSdSmetric}
\end{equation}
where $H$ is the Hubble rate (assumed to be constant),
related to the observed cosmological constant by
$\Lambda_\text{eff} = 3 H^2$. These field equations
(\ref{B1})--(\ref{B3}) then become long
expressions depending on the radial coordinate $r$ and
radial derivatives of the scalar field~(\ref{scalaransatz}).
However, we noticed that an extra hypothesis simplifies
them tremendously. In addition to the above assumptions
(\ref{eqSchwarzschildCoords}), (\ref{scalaransatz}) and
(\ref{eqSdSmetric}), we will also restrict to the case where
$X \equiv -(\varphi_\lambda)^2/M^2$ remains constant
everywhere, even in the vicinity of the massive body.
This means that we simply impose
\begin{equation}
X = X_c,
\label{Xconst}
\end{equation}
where $X_c = q^2$ is the constant cosmological value of $X$.
This choice is motivated by the exact Schwarzschild-de~Sitter
solution which has been obtained in the particular cases
of Refs.~\cite{Babichev:2013cya,Kobayashi:2014eva}.
All functions of $X$ entering Eqs.~(\ref{B1})--(\ref{B3}) then
obviously become constants, whose precise values are still
unknown, but which do not depend any longer on the radial
coordinate $r$. Moreover, since we have $X = e^{-\nu} q^2 -
e^{-\lambda}\varphi'^2/M^2$ in Schwarzschild coordinates
(\ref{eqSchwarzschildCoords}), where $\varphi' \equiv \partial_r
\varphi = \psi'$ denotes the radial derivative of the scalar
field (\ref{scalaransatz}), we may also replace any occurrence of
$\varphi'$ by the square root\footnote{Obviously, the r.h.s. of
Eq.~(\ref{phiPrime2}) needs to be positive for such an equation
to make sense, otherwise this would correspond to unstable
configurations. For example, such a situation takes place in
the model considered in~\cite{Babichev:2016jom} when the bare
cosmological constant is absent.} of
\begin{equation}
\varphi'^2 = e^{\lambda}\left(e^{-\nu}-1\right) M^2 X_c,
\label{phiPrime2}
\end{equation}
which is a \textit{known} function of $r$. Its radial derivative
also gives us the exact expression of $\varphi'' \equiv
\partial_r^2 \varphi$. Therefore, thanks to the greatly
simplifying hypothesis (\ref{Xconst}), the field equations
(\ref{B1})--(\ref{B3}) now become mere functions of $r$ alone,
involving some unknown constants depending of the functions
$f_n(X)$, $s_n(X)$ and their derivatives (with respect to $X$,
but also evaluated at $X= X_c$). Since these field equations must
be satisfied at any spacetime point, it is then straightforward
to extract from them some necessary conditions on the functions
$f_n(X)$ and $s_n(X)$. For instance, an expansion of
Eqs.~(\ref{B1})--(\ref{B3}) in powers of $(r-r_0)$ around any
radius $r_0$ (even $r_0 = 0$) suffices to prove that some
combinations of $f_n(X)$, $s_n(X)$ and their derivatives
\textit{must} vanish. After having derived such necessary
conditions, one may plug them back into
Eqs.~(\ref{B1})--(\ref{B3}) to check whether they also suffice.
If the field equations do not vanish identically, this means that
other conditions still need to be imposed. This procedure allowed
us to prove that the following conditions are necessary and
sufficient for the ansatz (\ref{eqSchwarzschildCoords}),
(\ref{scalaransatz}), (\ref{eqSdSmetric}) and (\ref{Xconst}) to
be consistent with all field equations (\ref{B1})--(\ref{B3}):
\begin{subequations}
\begin{eqnarray}
-X f_2 + 6\left(\frac{H}{M}\right)^2\left[X^2 f_4+ 2 X s_4\right]
&=& \frac{M_\text{Pl}^2}{M^4}\left(\Lambda_\text{bare} - 3 H^2\right),
\label{eqLambdaSdS}\\
\left[X f_2\right]' + 6\left(\frac{H}{M}\right)^2
\left[X^2 f_4+ 2 X s_4\right]' &=& 0,
\label{eqJSdS}\\
Xf_5+2 s_5 =0 \quad\text{and}\quad \left[Xf_5+2 s_5\right]' &=&0,
\label{eqf5SdS}\\
\left[X^{3/2}f_3\right]' &=&0,
\label{eqf3SdS}
\end{eqnarray}
\label{eqSdS}
\end{subequations}
where as before all functions $f_n$ and $s_n$ depend on $X$, and
a prime denotes differentiation with respect to $X$. This is
therefore a particular case of Eqs.~(\ref{eqCosmo}) that we
obtained in cosmology, which is not a surprise since the
asymptotic behavior of the present solution at large radii should
match with this cosmological solution. But we find here some
restrictions with respect to (\ref{eqCosmo}): The $f_3$ function
needs to be very precisely tuned at the cosmological value $X =
X_c$ (in order not to contribute to any background equation at
this precise value), while $f_5$ and $s_5$ should be related in a
specific way at this value of $X$ (again so that their sum
$Xf_5+2 s_5$ does not contribute to any background equation).

It should be underlined that this set of equations
(\ref{eqSdS}) only needs to be satisfied \textit{at} the
cosmological value $X = X_c$, and notably that
(\ref{eqf5SdS}) and (\ref{eqf3SdS}) should \textit{not}
be imposed for all $X$. Actually, if Eq.~(\ref{eqf3SdS}) were
satisfied for all $X$, then the Lagrangian $\mathcal{L}_{(3,0)}$,
Eq.~(\ref{eqL3}), would be a total derivative, as underlined in
Sec.~\ref{Sec3} above, and it would not contribute to any
observable. On the other hand, the sum
$\mathcal{L}_{(5,0)}+\mathcal{L}_{(5,1)}$, Eqs.~(\ref{eqL5}) and
(\ref{eqL51}), would not be a total derivative even if the two
conditions (\ref{eqf5SdS}) were imposed for all $X$. It just
happens that this combination does not contribute to the field
equations when imposing both spherical symmetry and $X =
\text{const.}$, as in the present section. Note that this
combination, satisfying conditions (\ref{eqf5SdS}) for all
$X$, is not the Horndeski one either, which would correspond to
$s'_5(X) = \frac{3}{4} f_5(X)$ (unless we are
in the limiting case $f_5\propto X^{-5/2}$ and $s_5\propto
X^{-3/2}$). The conditions (\ref{eqf5SdS}) and
(\ref{eqf3SdS}), at $X = X_c$, mean thus that $f_3$, $f_5$
and $s_5$ do not contribute to our Schwarzschild-de~Sitter
background solution, but they do change the behavior of
perturbations around this background, and they also change the
dynamics before the background reaches its equilibrium
configuration\footnote{\label{foot5}See our discussion below
Eqs. (6) for the combinations of Lagrangians avoiding the
presence of an extra ghost degree of freedom. This is notably so
when $f_5 = s_5 = 0$, whatever the other functions of $X$
defining the theory.}.

Equation~(\ref{eqLambdaSdS}) may also be rewritten as the
expression of the effective (observed) cosmological constant
$\Lambda_\text{eff} = 3 H^2$ in terms of $X = X_c$ and the bare
cosmological constant:
\begin{equation}
\Lambda_\text{eff}
=\frac{\Lambda_\text{bare}+ \frac{M^4}{M_\text{Pl}^2} X f_2}{1+
2\left(\frac{M}{M_\text{Pl}}\right)^2\left(X^2 f_4+ 2 X s_4\right)}.
\label{eqLambdaEff}
\end{equation}
This form underlines that $Xf_2$ acts as an additive constant to
$\Lambda_\text{bare}$ (recall that if $f_2 \propto 1/X$, then the
Lagrangian $\mathcal{L}_{(2,0)}$ would be another trivial bare
cosmological constant), whereas $X^2f_4$ and $Xs_4$ can be
understood as renormalization factors. However,
Eq.~(\ref{eqLambdaEff}) cannot be interpreted so directly because
these functions are anyway related via Eq.~(\ref{eqJSdS}), and
this explains notably why we found models in which
$\Lambda_\text{eff}$ is fully independent from
$\Lambda_\text{bare}$, at the end of Sec.~\ref{Sec3}. In the
realistic case where $3 H^2 = \Lambda_\text{eff} \ll
\Lambda_\text{bare}$, one may of course neglect $3H^2$ in the
r.h.s. of Eq.~(\ref{eqLambdaSdS}), and the added 1 in the
denominator of Eq.~(\ref{eqLambdaEff}) may thus be suppressed.

The conclusion of the present subsection is that a subclass of
beyond Horndeski theories does provide both cosmological
self-tuning, and a local metric around a spherical body which is
indistinguishable from GR plus a small $\Lambda_\text{eff}$. This
subclass depends on all six functions $f_n$ and $s_n$ defining
beyond Horndeski theories, Eqs.~(\ref{eqGeneralizedHorndeski}),
but they should satisfy the five relations~(\ref{eqSdS}) at the
background value $X = X_c$. Three of them, $f_2$, $f_4$ and
$s_4$, are responsible for the self-tuning, because
Eqs.~(\ref{eqLambdaSdS}) and (\ref{eqJSdS}) generically fix both
the values of $X_c$ and $3 H^2 = \Lambda_\text{eff}$ (see
Sec.~\ref{Sec3} for a discussion of the non-generic cases in
which one of those is not predicted). We shall call them the
``Three Graces''. The three other functions, $f_3$, $f_5$ and
$s_5$, play a passive r\^ole both for the cosmological background
and the local spherically symmetric solution, provided they
satisfy Eqs.~(\ref{eqf5SdS}) and (\ref{eqf3SdS}) at $X = X_c$.
Note however that the latter three ``stealth'' Lagrangians do
contribute to the dynamics of perturbations around our exact
solutions, and also to the time evolution of the Universe before
the equilibrium value $X = X_c$ is reached.

\subsection{Black hole solutions}
\label{Sec4B}
An interesting byproduct of the above Sec.~\ref{Sec4A}
is the existence of regular black hole
solutions.\footnote{Note that other black hole solutions
may also exist, while not respecting the hypotheses
of the present Section, notably the fact that we assume
$X = \text{const.}$} Indeed,
when conditions (\ref{eqSdS}) are imposed at $X = X_c$,
then the field equations admit the exact Schwarzschild-de~Sitter
solution (\ref{eqSchwarzschildCoords}) and (\ref{eqSdSmetric})
---~with an observed $\Lambda_\text{eff}$ much smaller than
$\Lambda_\text{bare}$. If no matter source is assumed for
$r>r_s$, the metric has thus the same form as that of
a general relativistic black hole.

Let us summarize here in which conditions such black holes
exist. We consider the most general beyond Horndeski Lagrangian
\begin{equation}
\mathcal{L}_{(2,0)}
+\mathcal{L}_{(3,0)}+\mathcal{L}_{(4,0)}+\mathcal{L}_{(4,1)}
+\mathcal{L}_{(5,0)}+\mathcal{L}_{(5,1)},
\end{equation}
defined by Eqs.~(\ref{eqGalileons}) and
(\ref{eqGeneralizedHorndeski}), but we require the following
three conditions at the cosmologically imposed value
$X = X_c = q^2$ of $X \equiv -(\varphi_\lambda)^2/M^2$:
\begin{subequations}
\label{cond35}
\begin{eqnarray}
X f_5+2 s_5 &=&0,\\
\left[Xf_5+2 s_5\right]' &=&0,\\
\left[X^{3/2}f_3\right]' &=&0.
\end{eqnarray}
\end{subequations}
On the other hand, the three other functions $f_2$, $f_4$
and $s_4$ are free, and they fix the value of $X_c$ from
Eqs.~(\ref{eqLambdaSdS}) and (\ref{eqJSdS}). Note that
conditions~(\ref{cond35}) do not need to be satisfied for
all values of $X$, but only at $X = X_c$. The metric
(\ref{eqSchwarzschildCoords}) reads then
\begin{equation}
e^\nu = e^{-\lambda} = 1-\frac{r_s}{r} - (H r)^2,
\label{eqSdSbis}
\end{equation}
with $\Lambda_\text{eff} = 3 H^2$ given by
Eq.~(\ref{eqLambdaEff}). The solution for the scalar
field is such that
\begin{equation}
X=q^2=\mathrm{const.},
\end{equation}
so that using the ansatz~(\ref{scalaransatz}),
one finds explicitly
\begin{equation}
\label{phiBH}
\varphi = q M t + \psi(r),
\end{equation}
with
\begin{equation}
\label{psiBH}
\psi' = - \frac{q M \sqrt{r_s/r + (H r)^2}}{1-r_s/r - (H r)^2}\,,
\end{equation}
the global minus sign being imposed by the matching
with the cosmological asymptotic behavior.
Although this last exact expression may be explicitly
integrated, let us quote here only the solution for a
negligible value of $H$:
\begin{equation}
\label{phiBH2}
\varphi = q M \left\{t- \left[2\sqrt{r_s r}
+ r_s \ln\left(\frac{\sqrt{r}
-\sqrt{r_s}}{\sqrt{r}+\sqrt{r_s}}\right)\right]
+ \mathcal{O}(H^2)\right\}.
\end{equation}

The regularity of such black-hole solutions is easy to understand.
First of all, since the metric is of the Schwarzschild-de~Sitter
form, it is clear that the backreaction of the scalar field on
the metric is everywhere finite, including both on the event and
the cosmological horizons. In fact, for these solutions, the
energy-momentum tensor of the scalar field (\ref{eqDefT})
takes precisely the form of a vacuum energy,
\begin{equation}
\frac{T_{\mu\nu}}{M_\text{Pl}^2}
= \left(\Lambda_\text{bare}- \Lambda_\text{eff}\right) g_{\mu\nu},
\label{eqScalarTmunu}
\end{equation}
as is obvious from our exact Schwarzschild-de~Sitter solution for
the metric. Concerning the regularity of the scalar field itself,
let us underline that the invariants $X$ and $J_\mu J^\mu$,
involving the derivatives of $\varphi$, are regular everywhere.
Indeed $X=q^2=\text{const.}$ by construction, while the current
$J^\mu$ actually fully vanishes for the present black hole
solutions. The reason is that by construction\footnote{For a
static and spherically symmetric black hole with the
time-dependence ansatz~(\ref{phiBH}),
Ref.~\cite{Babichev:2015rva} proved that $J^r=0$ follows from the
$0r$-Einstein equation.} $J^r=0$, while $J^0$ can be checked to
be also proportional to Eq.~(\ref{eqJSdS}) in the present case,
therefore the invariant $J^2 = J_\mu J^\mu$ vanishes everywhere.
The regularity of this norm of the current is an additional
condition which becomes important notably if matter is assumed
to be directly coupled to $J^\mu$. It is also one of the key
assumptions for the no-hair theorem for Galileons when the scalar
field is time-independent~\cite{Hui:2012qt} (contrary to our
present framework).

Although the scalar energy-momentum tensor takes the
very simple form~(\ref{eqScalarTmunu}) and the scalar
current $J^\mu$ vanishes, these solutions can anyway
be considered as ``hairy'' black holes, because of the
non-trivial configuration~(\ref{phiBH})--(\ref{psiBH}).
Note however that the notion of hair is not clearly defined
in the case of a time-dependent scalar field. Indeed,
a mere change of spatial hypersurface suffices to
create a spurious radial dependence. We may thus
consider a scalar (coordinate independent) quantity
which is constant in the cosmological background,
say $\Box\varphi = -3 H q M$, and show that it does
acquire a radial dependence in the above solution.
We find $\Box\varphi = -3 H q M -(3 q M r_s^2)/(8H^3 r^6)
+\mathcal{O}\left(r_s^3\right)$, which indicates the
existence of a scalar hair.

The black hole solutions presented here can be considered
as generalizations of the self-tuning solutions first found
for the Lagrangian containing the ``John'' term
$\mathcal{L}_{(4,1)}$~\cite{Babichev:2013cya}.
More specifically, this reference studied the model
$\mathcal{L}_{(2,0)}+\mathcal{L}_{(4,1)}+\Lambda_\text{bare}$,
with $f_2(X)=\text{const.}$ and $s_4(X)=\text{const.}$,
and showed that it admits an exact Schwarzschild-de~Sitter
solution describing a regular black hole.
Generalizations of this self-tuning solution have been found
later: Ref.~\cite{Babichev:2015qma} concluded that an
arbitrary $s_4(X)$ gives a similar solution, while the authors
of Ref.~\cite{Kobayashi:2014eva} considered the
Horndeski combination of $\mathcal{L}_{(4,0)}$ and
$\mathcal{L}_{(4,1)}$, i.e., with $s_4' = \frac{1}{4} f_4$
(see the recent review~\cite{Babichev:2016rlq}).

As a final remark, let us comment on our results above when the
$\mathcal{L}_{(3,0)}$ term is present. Equations~(\ref{cond35})
show that an exact Schwarzschild-de~Sitter solution exists only
for a particular class of functions $f_3$. This explains why
Ref.~\cite{Babichev:2016fbg} could not find simple solutions for
the theory with the simplest cubic Galileon, i.e., $f_3=1$. In
this case $f_3=1$, conditions~(\ref{cond35}) imply $q = X_c^{1/2}
= 0$, which is inconsistent with our hypothesis of a
time-dependent scalar field (\ref{phiBH}).

\section{Perturbations}
\label{Sec5}

\subsection{Backreaction of the scalar field}
\label{Sec5a}
The aim of the present Section is to go beyond the exact
solutions of the previous one, i.e., the Schwarzschild-de~Sitter
metric predicted in the ``Three Graces'' of Eqs.~(\ref{eqSdS})
when assuming $X = \text{const.}$ Can other models predict a
metric which is close enough to the Schwarzschild solution at
small radii, so that solar-system tests are passed? An obvious
first answer is to add small corrective terms to the Three
Graces, i.e., to assume that beyond the three functions $f_2$,
$f_4$ and $s_4$ satisfying \textit{approximately}
Eqs.~(\ref{eqLambdaSdS}) and (\ref{eqJSdS}), one adds
\textit{small enough} extra functions $f_3$, $f_5$ and $s_5$
which do not respect conditions~(\ref{eqf5SdS}) and
(\ref{eqf3SdS}).

But is it possible to pass solar-system tests in models which
differ significantly from the Three Graces~(\ref{eqSdS})? To
answer this question, we shall adopt here a perturbative
approach, whose spirit may be summarized as follows. In order to
pass solar-system tests, the metric should be close to the
Schwarzschild solution at small radii. We may thus assume that it
takes this approximate form to solve the scalar equation, and
then plug this scalar solution into the Einstein equations to
estimate its backreaction on the metric. Any contradiction will
prove that our approximations are not valid, i.e., that
solar-system tests cannot be passed.

We start from the most general field equations in spherical
symmetry, given in Eqs.~(\ref{B1})--(\ref{B3}) of Appendix
\ref{AppB}. Instead of expanding them around the Schwarzschild
solution, it is actually more convenient to assume that the
metric is almost flat, i.e., that the functions $\lambda$ and
$\nu$ entering (\ref{eqSchwarzschildCoords}) and their radial
derivatives are small with respect to 1. Our approximate Einstein
equations should therefore reproduce ultimately the linearized
behavior of the Schwarzschild solution. A mere linearization
would not be consistent for the scalar field itself, on the other
hand. Indeed, we know that nonlinear effects are crucial in
Galileon and Horndeski theories, for which a Vainshtein mechanism
generically exists at small radii. Our approximation scheme
should therefore take into account the powers of the radial
derivative $\varphi'$ entering the field equations. The only
hypothesis that we shall make is $|\varphi'| \ll |\dot\varphi| =
qM$, but we \textit{a priori} do not know the order of magnitude
of the second derivative $\varphi''$, and $\varphi'^2$ or higher
powers of $\varphi'$ are \textit{not} assumed to be negligible
with respect to $\lambda$ nor $\nu$. Finally, since the various
Lagrangians $\mathcal{L}_{(n,p)}$ involve different functions of
$X$, whose magnitude can be very different, we treat each of them
separately, without comparing the respective terms they generate.
But since we assume that none of the functions $f_n(X)$ nor
$s_n(X)$ involve large dimensionless parameters, we can consider
that $X f'_n \sim f_n$ for each of them separately.

It should be underlined that our hypothesis $|\varphi'| \ll
|\dot\varphi|$ might be problematic in the models where
$\dot\varphi = qM$ is predicted to be very small, since local
perturbations by the massive body might happen to be larger than
such a small cosmological background. Models involving negative
powers of $X$ might even yield to singularities in this case,
since $X$ may pass through zero, between its positive value at
cosmologically large distances and a negative one in the vicinity
of the massive body. In the following, we should thus trust our
perturbative treatment mainly when the cosmologically predicted
$|\dot\varphi|$ is not too small. However, we will see below that
in the most interesting subclass of models (the Three Graces),
the approximation $|\varphi'| \ll |\dot\varphi|$ is actually
justified even when $|\dot\varphi|$ is predicted to be extremely
small (with respect to the Planck mass), and even if the
Lagrangian involves negative powers of $X$. Our perturbative
treatment has thus clearly a wider range of application than
naively expected. In any case, one should keep in mind that the
conclusions of the present Section are valid only if our
hypothesis $|\varphi'| \ll |\dot\varphi|$ is satisfied.

The approximation scheme described above (linearization around a
flat metric, and for each different form of term, keeping the
lowest order in powers of $\varphi'/\dot\varphi$) then transforms
Eqs.~(\ref{B1})--(\ref{B3}) into the following form. It happens
that the time-time component of the Einstein equation
(\ref{eqEinstein}) can be easily integrated once with respect to
the radial coordinate, at this approximation. We thus quote below
its radial integral, multiplied by a global factor $M_\text{Pl}^2
r^2$ to simplify its expression. We then give the radial-radial
component of the linear combination
(\ref{eqCombLinEinsteinCurrent}) multiplied by a factor $M^4
M_\text{Pl}^2 r^2$, and finally the radial component of the
scalar current (\ref{eqDefCurrent}) multiplied by a factor $-2
M^2 r^2$:
\begin{subequations}
\begin{eqnarray}
M_\text{Pl}^2 r \left(\lambda
-\frac{1}{3}\, \Lambda_\text{bare} r^2\right)
+\frac{1}{3} M^4 X r^3 (f_2 + 2 X f_2')
+ M^2 X r^2 (3 f_3 + 2 X f_3') \varphi'\nonumber\\
+ 2 X r (5 f_4 + 2 X f_4') \varphi'^2
+ 2 X (7 f_5 + 2 X f_5') \varphi'^3/M^2\nonumber\\
+ 4 M^2 X r \lambda (s_4+2X s_4') + 8 r s_4 \varphi'^2
+ 4 X \lambda (3 s_5+ 2 X s_5') \varphi'
+ 8 s_5 \varphi'^3/M^2
&=& \frac{m}{4\pi},\quad\qquad
\label{eqG00}\\
\nonumber\\
M^4 M_\text{Pl}^2 \left(-\lambda + r \nu'
+\Lambda_\text{bare} r^2\right)
+ M^8 X r^2 f_2\nonumber\\
-2 M^4 X (\varphi' + 2 r \varphi'') \varphi' f_4
+ 6 \left(\nu'\varphi'^3-2M^2 X \varphi'' \right)
\varphi'^2 f_5\nonumber\\
+ 4 M^4\left[
M^2 X \lambda s_4
+ M^2 X r(s_4+ 2 X s_4') \nu'
- 2 s_4 \varphi'^2
+ 4 X r s_4' \varphi' \varphi''
\right]\nonumber\\
+ 4 M^2 X \left[
M^2 (3 s_5 + 2 X s_5')\nu'
+ 4 s_5' \varphi'\varphi''
\right]
\varphi' &=& 0,
\label{eqGrr}\\
\nonumber\\
4 M^4 r^2 (f_2 + X f_2')\varphi'
- M^2 r (3 f_3 + 2 X f_3')(M^2 X r \nu' - 4 \varphi'^2)\nonumber\\
- 4 M^2 X r \left[f_4 \lambda' +(5 f_4 + 2 X f_4')\nu'\right]\varphi'
+ 8 (2 f_4 + X f_4') \varphi'^3\nonumber\\
- 6 X \left[2 f_5 \lambda' + (7 f_5 + 2 X f_5')\nu'\right]\varphi'^2
+16 M^2 \left[X (\lambda+ r \lambda') s_4'
+ (\lambda - r \nu') s_4\right]\varphi'\nonumber\\
- 4 M^2 X \lambda(3 s_5 + 2 X s_5')\nu'
+ 8 (2 X s_5'\lambda' - 3 s_5 \nu')\varphi'^2
&=& 0.
\label{eqJr}
\end{eqnarray}
\label{eqPerturbations}
\end{subequations}
Here $X$ denotes the cosmological background $X_c = q^2$,
although we did not write its index to simplify the notation, and
all functions $f_n$ and $s_n$, as well as their derivatives, are
evaluated at $X_c$. Beware that the primes denote derivatives
with respect to the argument of the corresponding terms, i.e.,
$f_n' = df_n(X)/dX$ but $\varphi' = \partial_r\varphi$,
$\varphi'' = \partial_r^2\varphi$, $\lambda' = \partial_r\lambda$
and $\nu' = \partial_r\nu$.

For the same reason as in Sec.~\ref{Sec4} above, i.e., because we
assume there does not exist any direct matter-scalar coupling in
action (\ref{eqAction}), we know that $J^r = 0$ in the present
static and spherically symmetric situation, and this explains why
Eqs.~(\ref{eqGrr}) and (\ref{eqJr}) have vanishing right-hand
sides. On the other hand, the right-hand side $m/(4\pi)$ of
Eq.~(\ref{eqG00}) is imposed by the matching of this equation
with the interior of the massive body, whose total mass is
denoted $m$. The matter contribution to $T_{00}$ is indeed the
matter density $\rho$ (at this order of approximation), and we
have $\int \rho\, r^2 dr = m /(4\pi)$.

The analysis of Eqs.~(\ref{eqPerturbations}) can be decomposed in
three different cases, depending on which beyond Horndeski
Lagrangians dominate at small distances. It is indeed expected
that only one of them dominates locally [although it may happen
that several of them simultaneously dominate, when their
functions $f_n(X)$ or $s_n(X)$ are tuned to obtain such a
behavior]. For instance, the local domination of the
$\mathcal{L}_{(3,0)}$ term would be related to the well-known
Vainshtein mechanism. It should thus be kept in mind that the
cosmological background (and notably the predicted value of
$\Lambda_\text{eff}$) may not depend on the same set of terms as
those which dominate at small distances. We will thus in general
treat the local equations~(\ref{eqPerturbations}) without
assuming that the same functions are responsible for the
cosmological background.

The first case one may consider is when $f_2$, $f_4$ and/or
$s_4$ dominate at small distances. This corresponds to the
Three Graces, and our results of Sec.~\ref{Sec4} show that
an exact Schwarzschild-de~Sitter solution is then possible. It is
thus obvious that the linearized equations (\ref{eqPerturbations})
are also consistent with a local Schwarzschild metric, and it is
not necessary to check again so. We will see in Sec.~\ref{Sec5b}
below that these linearized equations (\ref{eqPerturbations}) are
nevertheless useful in this Three Graces case too, to study the
renormalization of Newton's constant.

The second case we consider is when $f_3$ happens to dominate at
small distances. Then Eq.~(\ref{eqJr}) tells us that either
$3 f_3 + 2 X f_3' = 0$ or $4 \varphi'^2 = M^2 X r \nu'$. But if
$3 f_3 + 2 X f_3' = 0$, then $f_3$ is fully passive (cf.~our
cosmological discussion in Sec.~\ref{Sec3}, and the fact that
$\mathcal{L}_{(3,0)}$ is a total derivative when such a condition
is imposed for all $X$), and it cannot dominate at small
distances. Therefore, we must have $4 \varphi'^2 =\dot\varphi^2 r
\nu'$, and if we assume that the metric is approximately of the
Schwarzschild form (to pass solar-system tests), i.e., $\nu
\approx -r_s/r$, we thus get $4 \varphi'^2 =\dot\varphi^2 r_s/r$.
Plugging this back into Eq.~(\ref{eqG00}), we find that the
backreaction of the scalar is
\begin{equation}
M^2 X r^2 (3 f_3 + 2 X f_3') \varphi'
= \frac{1}{2}\,\dot\varphi^3 (3 f_3 + 2 X f_3')\sqrt{r_s r^3}.
\label{eqBackreaction}
\end{equation}
This is to be compared to the r.h.s. of Eq.~(\ref{eqG00}), namely
$m/(4\pi)$. Depending on which Lagrangians determine the
cosmological evolution, it may happen that this backreaction is
negligible, and therefore that solar-system tests can be passed
[although this situation would need a well-chosen function
$f_3(X)$]. However, this is not the case when
$\mathcal{L}_{(3,0)}$ also contributes significantly to
cosmology. Let us illustrate so on the simple example of
$\mathcal{L}_{(2,0)}+ \mathcal{L}_{(3,0)}$ with monomials $f_2 =
k_2 X^\alpha$ and $f_3 = k_3 X^\beta$. Then the cosmological
equations (\ref{eqCosmo}) imply that we \textit{always} have
$\dot\varphi^3 f_3 \sim M_\text{Pl}^2 \Lambda_\text{bare}/H$, up
to $\mathcal{O}(1)$ factors, whatever the exponents $\alpha$ and
$\beta$ entering the monomials. Therefore, the
backreaction~(\ref{eqBackreaction}) is always of order
$M_\text{Pl}^2 \Lambda_\text{bare}\sqrt{r_s r^3}/H \sim
\left[(\Lambda_\text{bare}/H^2) (\Lambda_\text{bare} r_s^2)
(r/r_s)^3\right]^{1/2} m$, which is much larger than $m$ because
the term within the square brackets is a product of three large
numbers.\footnote{If $M_\text{Pl}^2 \Lambda_\text{bare}$ is
assumed to take the smallest possible theoretical prediction,
namely $|\rho_\text{QCD}|\sim 10^{-2} \text{GeV}^4$, then
$\Lambda_\text{bare} r_s^2$ would actually be of order
$\mathcal{O}(1)$ for the Schwarzschild radius of the Sun, but
this is anyway multiplied by the large factors
$\Lambda_\text{bare}/H^2$ and $(r/r_s)^3$. Let us also mention
that Newton's constant $G$ is not renormalized in the present
model, contrary to those discussed in Sec.~\ref{Sec5b} below, and
therefore that it is legitimate to identify here $r_s$ and $2 Gm
= m /(4\pi M_\text{Pl}^2)$.} In conclusion, in this simple
$\mathcal{L}_{(2,0)}+ \mathcal{L}_{(3,0)}$ model, the metric
cannot be close to the Schwarzschild solution, and solar-system
tests are not passed. The only ways out are either that the
contribution of $\mathcal{L}_{(3,0)}$ is negligible in the
cosmological equations (\ref{eqCosmo}), so that $\dot\varphi$ is
actually unrelated to $f_3$ and the backreaction $\propto
\dot\varphi^3 f_3$ can be small enough, or that other Lagrangians
than $\mathcal{L}_{(3,0)}$ dominate at small distances, which
depends on the functions $f_n(X)$ entering them.

The third and final case is when $f_5$ and/or $s_5$ dominate at
small distances. If we assume an approximate Schwarzschild
metric, then Eq.~(\ref{eqJr}) implies
\begin{equation}
\varphi'^2 = \frac{-2 M^2 X r_s (3 s_5 + 2 X s_5')}{r
\left[ 3 X (5 f_5 + 2 X f_5') + 4 (3 s_5 + 2 X s_5') \right]}.
\label{eqPhiPrimeL5}
\end{equation}
Note that $f_5$ alone (with $s_5 = 0$) is not allowed to dominate
in the vicinity of the massive body, otherwise its contribution
would violate Eq.~(\ref{eqJr}). [The only way out would be to
impose $(X^{5/2} f_5)' = 0$, in which case it would actually not
dominate locally.] In fact, $s_5$ alone (with $f_5 = 0$) is not
allowed either to dominate locally, otherwise Eq.~(\ref{eqPhiPrimeL5})
would give a negative $\varphi'^2$. We should thus assume that
both $f_5$ and $s_5$ dominate simultaneously.
Plugging the expression (\ref{eqPhiPrimeL5}) of $\varphi'^2$ into
Eq.~(\ref{eqG00}) gives us the backreaction of the scalar field
on the metric
\begin{equation}
-\frac{8}{M^2}\left[X (2 f_5 + X f_5')
+ 2 (s_5 + X s_5')\right] \varphi'^3,
\label{eqL5BackReaction}
\end{equation}
[with $\varphi'$ still given by Eq.~(\ref{eqPhiPrimeL5})], which
is again to be compared to $m/(4\pi)$, i.e., the r.h.s. of
Eq.~(\ref{eqG00}). Similarly to the case of $\mathcal{L}_{(3,0)}$
above, it may happen that this backreaction is negligible if the
cosmology is determined by other Lagrangians than
$\mathcal{L}_{(5,0)}$ or $\mathcal{L}_{(5,1)}$, although $f_5$
and $s_5$ are assumed to dominate at small distances [this would
also need some well-chosen functions $f_5(X)$ and $s_5(X)$]. But
if these Lagrangians do contribute significantly to the
cosmological background, then we face again the same difficulty
as for $\mathcal{L}_{(3,0)}$: The backreaction of the scalar
field is much larger than the central source $m/(4\pi)$. This can
be illustrated on the simple example of $\mathcal{L}_{(2,0)}+
\mathcal{L}_{(5,0)}+ \mathcal{L}_{(5,1)}$ with monomials $f_2 =
k_2 X^\alpha$, $f_5 = k_5 X^\beta$ and $s_5 = \kappa_5 X^\gamma$.
Then one finds that the backreaction (\ref{eqL5BackReaction}) is
always of order $M_\text{Pl}^2 \Lambda_\text{bare} H^{-3}
(r_s/r)^{3/2} \sim \left[(\Lambda_\text{bare}/H^2)^3
(\Lambda_\text{bare} r_s^2) (H r)^{-6}\right]^{1/4} m$, which is
much larger than $m$ because the term within the square brackets
is a product of three large numbers.

However, even when such an \textit{a priori} large backreaction
is expected, there still exists one possibility to pass
solar-system tests. It suffices that
\begin{equation}
X (2 f_5 + X f_5') + 2 (s_5 + X s_5') = 0,
\label{eqFS5}
\end{equation}
since this factor multiplies the backreaction
(\ref{eqL5BackReaction}). On the other hand, note that it
would not be possible to impose $(3 s_5 + 2 X s_5') = 0$
[cf. our limiting case discussed in Sec.~\ref{Sec3} below
Eqs.~(\ref{eqCosmo})], although this would also give a vanishing
backreaction. Indeed, this would correspond to $\varphi' = 0$ in
Eq.~(\ref{eqPhiPrimeL5}), in contradiction with our hypothesis
that $f_5$ and $s_5$ dominate the local physics of $\varphi$. But
condition (\ref{eqFS5}) may be imposed without any inconsistency
nor obtaining a trivial model. One can also check that the
dominant ($f_5$ and $s_5$) terms of the second Einstein equation
(\ref{eqGrr}) also vanish when this condition is assumed. The two
Einstein equations therefore reduce to those of general
relativity when condition~(\ref{eqFS5}) is imposed, and
Schwarzschild solution is recovered at small distances.

Note that Eq.~(\ref{eqFS5}) would be a consequence of the two
conditions (\ref{eqf5SdS}) we found to get our exact
solution of Sec.~\ref{Sec4}, but it does not suffice to imply
both of them. In the present approximation scheme, we find thus
that less constraints are needed to predict a Schwarzschild
solution. It is probable that a higher-order analysis, taking
into account first post-Newtonian terms in the $g_{00}$ component
of the metric [which are of order $(r_s/r)^2$], would imply a
second condition, and that we would then recover the two of
Eqs.~(\ref{eqf5SdS}). But at the present linear order in
$r_s$, the only conclusion we can draw is that the combination of
Lagrangians $\mathcal{L}_{(5,0)}+\mathcal{L}_{(5,1)}$ only needs
to satisfy the single condition~(\ref{eqFS5}) to be consistent
with a Schwarzschild metric when they dominate locally, whatever
the cosmological behavior [which may depend on other Lagrangians
$\mathcal{L}_{(n,p)}$] and even if it yields very large factors
multiplying the $f_5$ and $s_5$ terms in the local equations.

In conclusion, when $\mathcal{L}_{(5,0)}+\mathcal{L}_{(5,1)}$
dominate the behavior of $\varphi$ in the vicinity of a massive
body, there are two ways to pass solar-system tests. The first
one is similar to the case of $\mathcal{L}_{(3,0)}$ above, namely
when the cosmological evolution, depending on other Lagrangians
$\mathcal{L}_{(n,p)}$, is such that the backreaction
(\ref{eqL5BackReaction}) is small enough with respect to the mass
$m$ of the body (but this needs some well-chosen functions $f_5$
and $s_5$). The second possibility is to choose a model
satisfying condition~(\ref{eqFS5}), which is a subset of
Eqs.~(\ref{eqf5SdS}) found for the exact solutions of
Sec.~\ref{Sec4}. Then the scalar field does not backreact at all
on the metric (when $f_5$ and $s_5$ locally dominate) whatever
the cosmological solution.

\subsection{Renormalization of Newton's constant}
\label{Sec5b}
Although the quantity $2 Gm = m /(4\pi M_\text{Pl}^2)$ entering
Eq.~(\ref{eqG00}) would be called the Schwarzschild radius of the
body in standard general relativity, one should keep in mind that
in the present class of theories, this is \textit{not} the
coefficient entering the possible $\mathcal{O}(1/r)$ terms in
$-g_{00}$ and $g_{rr}$. Indeed, the scalar field also contributes
crucially to the behavior of the metric, and one does not even
predict a Newtonian potential $\propto 1/r$ in most models. Even
in the exact solutions of Sec.~\ref{Sec4} where the metric happens
to take the Schwarzschild-de~Sitter form,
Eqs.~(\ref{eqSchwarzschildCoords}) and (\ref{eqSdSmetric}), the
Schwarzschild radius $r_s$ entering its expression does
generically differ from $2 G m$.

Let us indeed consider the particular case in which only $f_2$,
$f_4$ and $s_4$ dominate at small enough distances, i.e., the
Three Graces of Eqs.~(\ref{eqSdS}). Let us also assume
that $X = q^2 = \text{const.}$, like in Sec.~\ref{Sec4}, which
implies
\begin{equation}
\varphi'^2 = e^\lambda (e^{-\nu}-1) M^2 q^2
= M^2 q^2 r_s/r + \mathcal{O}(r_s^2/r^2) +
\mathcal{O}(\Lambda_\text{eff} r^2).
\label{eqPhiPrime}
\end{equation}
Note that this means we always have $\varphi'^2 \ll \dot\varphi^2
= M^2 q^2$, i.e., the condition we assumed to make the expansions
of Sec.~\ref{Sec5a}, even in the cases where $|\dot\varphi|$ will
be predicted to be extremely small with respect to the Planck
mass.

Then, the constant contributions to Eq.~(\ref{eqG00}) (neglecting
those $\propto r^3$ which dominate at larger distances) imply
\begin{equation}
r_s = \frac{2Gm}{1 + 4 \left(\frac{M}{M_\text{Pl}}\right)^2 X^{1/2}
\left[X^{5/2} f_4 + 2 X^{3/2} s_4\right]'}\, ,
\label{eqRs}
\end{equation}
where the prime denotes derivation with respect to $X$. This is
equivalent to a renormalization of Newton's constant $G$ by the
denominator of (\ref{eqRs}). This renormalization does depend on
the cosmological background via $X$, but note that it is body
independent. In other words, it cannot be distinguished from
general relativity by local experiments, even by equivalence
principle tests involving three bodies or more. It suffices that
the ratio of the bare gravitational constant $G$ and the
denominator of (\ref{eqRs}) take the experimental value of
Newton's constant. [Note that we are talking here only of the
non-observable effect caused by this renormalization of $G$.
There may exist other deviations from GR in three-body systems,
for instance preferred-frame effects, that we do not discuss in
the present paper.]

In the realistic situation where the observed
$\Lambda_\text{eff}$ is much smaller than $\Lambda_\text{bare}$,
the added $1$ in the denominator of Eq.~(\ref{eqRs}) is
generically negligible. It is indeed dominated by the second
term involving functions of $X$, which is of the same order
of magnitude as those entering the cosmological equations
(\ref{eqCosmo}), or more precisely Eqs.~(\ref{eqLambdaSdS})
and (\ref{eqJSdS}) in the present Three Graces. Combining
these equations with (\ref{eqRs}), we thus generically predict
\begin{equation}
\left(M_\text{Pl}^\text{bare}\right)^2 \Lambda_\text{bare}
\sim \left(M_\text{Pl}^\text{eff}\right)^2 \Lambda_\text{eff},
\label{eqMPlbare}
\end{equation}
up to $\mathcal{O}(1)$ numerical factors, where
$M_\text{Pl}^\text{bare}$ means our previous notation
$M_\text{Pl}$, while $M_\text{Pl}^\text{eff}$ is the numerical
value corresponding to the actually measured Newton's constant.
For instance, in example~(\ref{eqExample1}), one gets $5
\left(M_\text{Pl}^\text{bare}\right)^2 \Lambda_\text{bare} = 3
\left(M_\text{Pl}^\text{eff}\right)^2 \Lambda_\text{eff}$, while
example~(\ref{eqExample2}) gives $3
\left(M_\text{Pl}^\text{bare}\right)^2 \Lambda_\text{bare} =
\left(M_\text{Pl}^\text{eff}\right)^2 \Lambda_\text{eff}$. Let us
recall that quantum field theory should give the value of the
vacuum \textit{energy density} from the matter action
$S_\text{matter}$ of Eq.~(\ref{eqAction}). Although we decide to
write it as a product $\left(M_\text{Pl}^\text{bare}\right)^2
\Lambda_\text{bare}$ in this action, it is \textit{a priori}
unrelated to Newton's constant nor to the observed accelerated
expansion of the Universe. The cosmological constant problem is
precisely that the measured values of $G$ (e.g. by Cavendish
experiments) and of the cosmological constant (e.g. from type-Ia
supernovae data) give a product
$\left(M_\text{Pl}^\text{eff}\right)^2 \Lambda_\text{eff}$ much
too small, by many orders of magnitude, with respect to the
expected vacuum energy density
$\left(M_\text{Pl}^\text{bare}\right)^2 \Lambda_\text{bare}$. In
the present scenario, Eq.~(\ref{eqMPlbare}) implies thus that the
cosmological constant problem is not solved at all, and not even
alleviated: The observable quantity
$\left(M_\text{Pl}^\text{eff}\right)^2 \Lambda_\text{eff}$
actually keeps the same order of magnitude as the huge bare
vacuum energy density!

However, the generic behavior (\ref{eqMPlbare}) is no longer valid
if the denominator of Eq.~(\ref{eqRs}) is not large, and this can
happen without any fine tuning if the functions $f_4$ and $s_4$
are chosen so that
\begin{equation}
\left[X^{5/2} f_4 + 2 X^{3/2} s_4\right]' = 0,
\label{eqNoRenorm}
\end{equation}
at $X =X_c $. This condition obviously reduces the space of
allowed models, but it does not need any large nor small
dimensionless number to be imposed. The combination $X^{5/2} f_4
+ 2 X^{3/2} s_4$ itself must not vanish, otherwise the field
equations (\ref{eqLambdaSdS}) and (\ref{eqJSdS}) cannot be
satisfied (unless $f_2 \propto 1/X$, meaning that
$\mathcal{L}_{(2,0)}$, Eq.~(\ref{eqL2}), is a second bare
cosmological constant). We must thus choose
\begin{equation}
\label{extraCond}
X^{5/2} f_4 + 2 X^{3/2} s_4 = \text{const.}
\end{equation}
Many possibilities exist in which
$f_4$ and $s_4$ almost compensate each other apart from this
constant, but they all give the same physics both in the
cosmological framework of Sec.~\ref{Sec4} and in our exact
solutions for spherical symmetry of Sec.~\ref{Sec5}. It suffices
thus to consider the simplest cases of $f_4 = k_4 X^{-5/2}$
and/or $s_4 = \kappa_4 X^{-3/2}$, where $k_4$ and $\kappa_4$ are
dimensionless constants of order~1. Then, Eq.~(\ref{eqRs})
implies that we have strictly $M_\text{Pl}^\text{bare} =
M_\text{Pl}^\text{eff}$ in this subclass of the Three Graces.
In conclusion, the extra condition (\ref{eqNoRenorm}), added
to Eqs.~(\ref{eqSdS}), now allows us to predict a small observed
$\Lambda_\text{eff}$ while keeping the Planck mass
unrenormalized, so that the observed vacuum energy density
$\left(M_\text{Pl}^\text{bare}\right)^2 \Lambda_\text{eff}$ may
be as small as wished.

Note that the six conditions (\ref{eqSdS}) and (\ref{eqNoRenorm})
only need to be satisfied at one value of $X = X_c$. Therefore,
there still remain six free functions, which do contribute to the
evolution of the Universe before it reaches its equilibrium at
$X = X_c$, as well as to the dynamics in generic non-symmetric
situations or for perturbations around a spherically symmetric
solution\footnote{See again our discussion below Eqs. (6) for the
combinations of Lagrangians avoiding the presence of an extra
ghost degree of freedom.}. However, the only physically relevant
terms of the action, for our exact Schwarzschild-de~Sitter
background, are just a free $f_2(X)$ and $f_4 = k_4 X^{-5/2}$
and/or $s_4 = \kappa_4 X^{-3/2}$. All the other functions,
including some non-trivial contributions to $f_4$ and $s_4$ which
cancel in the combination (\ref{eqNoRenorm}), are passive for
this solution, i.e., do not enter the result.

An example of a model satisfying all conditions (\ref{eqSdS}) and
(\ref{eqNoRenorm}) is given in Eqs.~(\ref{eqExample3}) above.
Since both $f_4 = k_4 X^{-5/2}$ and $s_4 = \kappa_4 X^{-3/2}$ are
allowed, it is also possible to use the Horndeski combination,
such that $F_4 = 0$ in Eq.~(\ref{eqF4}). Then all field equations
involve at most second derivatives, which simplifies their
analysis (although the third derivatives of generalized Horndeski
models with $F_4 \neq 0$ do not generate an extra degree of
freedom, as recalled in Sec.~\ref{Sec2}). In the present case,
$F_4 = 0$ implies $k_4 = -6 \kappa_4$, and this corresponds to
$G_4(-M^2 X) = -2 M^2 \kappa_4 X^{-1/2}$ in Eq.~(\ref{eqG4}). Let
us choose $k_2 = \kappa_4 = -1$ to simplify. Then the specific
model $f_2 = s_4 = -X^{-3/2}$ and $f_4 = 6 X^{-5/2}$ is in the
Horndeski class, and does not predict any renormalization of
Newton's constant. It also predicts that the observed Hubble rate
$H = M/(2\sqrt{6})$ is fully independent from the bare vacuum
energy density $M_\text{Pl}^2 \Lambda_\text{bare}$ involved in
action~(\ref{eqAction}), and therefore does not change even after
phase transitions possibly modifying this vacuum energy. On the
other hand, this means that the Hubble scale $H$ needs to be
introduced by hand in the action via the mass scale $M$,
therefore there still exist some fine-tuning in such a model,
although it concerns the mass scale entering the action of a
scalar field instead of the vacuum energy itself. A better model
may be for instance $f_2 = -X^{-5/4}$, $f_4 = 6 X^{-5/2}$ and
$s_4 = -X^{-3/2}$, which is still in the Horndeski class and does
not predict any renormalization of Newton's constant, but which
now needs $M = (32 M_\text{Pl}^2 \Lambda_\text{bare} H^2)^{1/6}$.
In such a case, the mass scale $M$ introduced in the action is
thus intermediate\footnote{If the vacuum energy density
$M_\text{Pl}^2 \Lambda_\text{bare}$ is of order $c^7/(\hbar G^2)$,
then this corresponds to $M \sim 100\ \text{MeV}/c^2$, similar
to usual elementary particle masses.} between the huge Planck
mass and the tiny Hubble rate.

\section{Conclusions}
\label{Concl}
In this paper, we studied self-tuning in all shift-symmetric
beyond Horndeski theories. Our goal is two-fold. First, we
demonstrate that the theory does provide a mechanism to almost
fully screen a very large bare cosmological constant entering the
action, leaving a small effective (observable) one consistent
with the present accelerated expansion of the Universe. Second,
we select a subclass of beyond Horndeski theories which not only
provide such a self-tuning of the cosmological constant, but also
do not contradict Solar system tests.

Our starting point is the beyond Horndeski
action~(\ref{eqAction}) with only two mass scales in the action,
the Planck mass $M_\text{Pl}$ and an extra scale $M$. The theory
contains six arbitrary functions, which specify the different
possible kinetic terms of the scalar field, see
Eqs.~(\ref{eqGeneralizedHorndeski}). We then progressively reduce
the space of allowed models by imposing different physical
requirements.

First we show that self-tuning is possible for a generic
combination of beyond Horndeski Lagrangians, provided that the
parameter $M$ is adjusted to predict $\Lambda_\text{eff} \ll
\Lambda_\text{bare}$. At this level all the six functions of the
theory are still allowed, the only constraint being on the
magnitude of $M$ ---~which may be either large or small with
respect to $\left(M_\text{Pl}^2
\Lambda_\text{bare}\right)^{1/4}$, depending on the model, but
not of the same order of magnitude.

As a second step, we ask that the Schwarzschild-de~Sitter (SdS)
metric is a solution of the theory. This is a sufficient
condition to satisfy (basic) Solar system tests of gravity. We
find that an exact SdS solution does exist when the scalar field
is such that $\varphi_\lambda \varphi^\lambda = \text{const.}$,
provided the five conditions~(\ref{eqSdS}) are satisfied.
Although the six functions still play a r\^ole before the
Universe reaches this solution, as well as for the dynamics of
perturbations around this solution, the conditions~(\ref{eqSdS})
effectively switch off three of them from the cosmological
de~Sitter evolution (and the SdS solution), making them passive
(or ``stealth''), so that the SdS solution does not feel them.
The other three functions are $f_2$, $f_4$ and $s_4$, that we
call the ``Three Graces''. They are responsible for the resulting
cosmological and SdS solution.

As a by-product of the above study, we found a class of regular
black hole solutions, which can be considered as generalization
of the self-tuning solutions found
in~\cite{Babichev:2013cya,Babichev:2015qma,Kobayashi:2014eva}.
Namely, beyond Horndeski theory satisfying
conditions~(\ref{cond35}) at $X=q^2$, where $\dot \varphi = q M$
is the cosmological value of the scalar field time derivative,
allows for self-tuning Schwarzschild-de~Sitter black hole
solutions with metric~(\ref{eqSdSbis}) and the non-trivial scalar
field~(\ref{phiBH}), (\ref{psiBH}).

Then we study perturbative corrections to the above solutions,
allowing slightly non-SdS solutions. Doing so, we relax the above
strict condition that the local solution must be of the exact
Schwarzschild form. This allows us to take into account small
deviations from GR which might not be observable with the present
precision of local gravity tests. We find that in addition to the
above Three Graces, the three other beyond Horndeski Lagrangians
may give a small enough backreaction of the scalar field on the
metric, notably when the local physics and the asymptotic
cosmological behavior are not dominated by the same terms of the
Lagrangian. On the other hand, when the same terms play a
significant r\^ole both at small and large distances, the scalar
backreaction is generically so large that Solar-system tests
cannot be passed. There remains however one interesting subclass
of models, satisfying condition~(\ref{eqFS5}), such that the
deviations from the local Schwarzschild solution are small
enough, even when the corresponding Lagrangians contribute
significantly both at large and small distances. This
condition~(\ref{eqFS5}) is a subset of the two (\ref{eqf5SdS})
we found when imposing an exact SdS solution.

It turns out, however, that when we take into account the
renormalization of Newton's constant $G$, which naturally happens
for a time-dependent scalar field in the theory under
consideration, the cosmological problem is \textit{not} solved.
This happens because the effective vacuum energy density has
approximately the same value as the bare vacuum energy density,
the two effects ---~effective decreasing of the cosmological
constant and the effective increasing of the Planck mass~---
almost compensating each others, see Eq.~(\ref{eqMPlbare}). In
order to solve the cosmological constant problem, while taking
into account the renormalization of $M_\text{Pl}$, we need to
impose the extra condition~(\ref{eqNoRenorm}), in addition
to~(\ref{eqSdS}). At this stage, we find that two out of the
three functions entering the Three Graces must be very specific
power laws, and there only remains one free function, $f_2(X)$,
defining this subclass of allowed models.

To summarize, we found that the subclass of beyond Horndeski
theory satisfying the six conditions~(\ref{eqSdS}) and
(\ref{eqNoRenorm}) does solve the big cosmological constant
problem, without any obvious contradiction with Solar system
gravity tests.

More detailed analysis of Solar-system constraints is left for
future work. Indeed we showed that we can choose the beyond
Horndeski action such that the theory admits an exact SdS
solution. However, this does not necessarily mean that all local
gravity tests are passed. Indeed, perturbations of planets (which
are not included in our analysis) may give deviations from GR.
For instance, the Nordtvedt effect, which tests the strong
equivalence principle, would need to be studied in the present
framework. It is tightly constrained by the \textit{three}-body
system Earth-Moon-Sun. The physics of the \textit{interior}
of stars may also be a way to additionally constrain these
theories, notably because there exist couplings to the
derivatives of the matter density in beyond Horndeski
theories~\cite{Kobayashi:2014ida,Babichev:2016jom}.

Finally, the stability of the above SdS solutions is yet to be
understood. We do know that some ghost or gradient
instabilities exist in some models (for instance for
$\mathcal{L}_{(2,0)}+\mathcal{L}_{(3,0)}$ in this self-tuning
scenario), but this needs to be studied for the more promising
Three Graces. We also leave this study for future work.

\section*{Acknowledgments}
We wish to thank Christos Charmousis for enlightening
discussions. E.B. was supported in part by the research program
``Programme national de cosmologie et galaxies'' of the
CNRS/INSU, France, and Russian Foundation for Basic Research
Grant No. RFBR 15-02-05038.

\appendix
\section{Partial integration of the beyond-Horndeski Lagrangians}
\label{AppA}
The Lagrangians (\ref{eqGeneralizedHorndeski}) may be integrated
by parts to be rewritten as follows:
\begin{eqnarray}
\mathcal{L}_{(3,0)} &=&
-M^2\left[X f_3(X) +\frac{1}{2}\int f_3(X)dX\right]
\Box\varphi +\text{tot. div.},
\label{eqPartialIntegrations3}\\
\mathcal{L}_{(4,0)} + \mathcal{L}_{(4,1)} &=&
-2 M^2 X s_4(X) R\nonumber\\
&&-\left[Xf_4(X) + \int f_4(X) dX\right]
\Bigl[\left(\Box\varphi\right)^2
- \varphi_{\mu\nu}\varphi^{\mu\nu}\Bigr]\nonumber\\
&&+ \left[ \int f_4(X) dX
- 4 s_4(X)\right] R^{\mu\nu} \varphi_\mu\varphi_\nu
+\text{tot. div.},
\label{eqPartialIntegrations4}\\
\mathcal{L}_{(5,0)} + \mathcal{L}_{(5,1)} &=&
-\left[\frac{3}{2}\int\!\!\!\!\int
f_5(X) dX dX - 4 X s_5(X)-4\int s_5(X) dX\right]
G^{\mu\nu} \varphi_{\mu\nu}\nonumber\\
&&-\frac{1}{2M^2}\left[2 X f_5(X) + 3\int f_5(X)dX\right]
\Bigl[\left(\Box \varphi\right)^3
- 3\, \Box\varphi\, \varphi_{\mu\nu}\varphi^{\mu\nu}
+ 2\, \varphi_{\mu\nu}\varphi^{\nu\rho}
\varphi_\rho^{\hphantom{\rho}\mu}\Bigr]
\nonumber\\
&&-\frac{1}{2M^2}\left[3\!\!\int\!\! f_5(X)dX - 4 s_5(X)\right]
\nonumber\\
&&\times
\Bigl[R\, \varphi^\mu \varphi_{\mu\nu} \varphi^\nu
-2\, \Box\varphi\, R^{\mu\nu}\varphi_\mu\varphi_\nu
+ 2 R^{\mu\nu\rho\sigma}\varphi_\mu
\varphi_\rho \varphi_{\nu\sigma}\Bigr]
+\text{tot. div.}
\label{eqPartialIntegrations5}
\end{eqnarray}
These expressions ease the translation of our notation
(\ref{eqGeneralizedHorndeski}) in terms of the functions
$G_n$, $F_n$, $A_n$ and $B_n$ used in the literature,
and explicitly given in Eqs.~(\ref{eqG}) and (\ref{eqA}) above.
Note that the first term of Eq.~(\ref{eqPartialIntegrations5})
involves a double primitive of $f_5(X)$, i.e., a primitive of
the single integral $\int\! f_5(X)dX$ entering other terms.

\section{Field equations in a static and spherically symmetric
situation}
\label{AppB}
We give below the field equations of the most general
shift-symmetric beyond Horndeski theory (\ref{eqAction}) when the
metric is assumed to be static and spherically symmetric, in
Schwarzschild coordinates~(\ref{eqSchwarzschildCoords}), while
imposing that the scalar field has the linear time dependence
(\ref{scalaransatz}). These equations are used in
Secs.~\ref{Sec4} and \ref{Sec5}, in which we first simplify them
considerably by assuming $X \equiv -(\varphi_\lambda)^2/M^2 =
\text{const.}$, and then linearize them around a flat metric for
$|\varphi'| \ll |\dot\varphi|$.

Let us display first the time-time component of the Einstein
equation (\ref{eqEinstein}), globally multiplied by a factor
$M^6 M_\text{Pl}^2 r^2$:
\begin{eqnarray}
&&M^8 r^2 \left[M^2 q^2 \left(f_2+2 X f_2'\right)
+e^{\nu-\lambda} \varphi'^2 f_2 \right]\nonumber\\
&&+\frac{1}{2} e^{-2 \lambda} M^6
r \left(3 f_3+2 X f_3'\right) \left[e^{\nu} r \varphi'^2
\left(\lambda' \varphi'-2 \varphi''\right)
+e^{\lambda} M^2 q^2 \left(\varphi'
\left(4-r \lambda'\right)+2 r \varphi''\right)\right]\nonumber\\
&&+e^{-3 \lambda}
\Bigl[2 e^{\nu} M^4 \varphi'^3 \left\{\varphi' \left(r \lambda'
\left(5 f_4+2 X f_4'\right)-f_4\right)
-4 r \varphi'' \left(2 f_4+X f_4'\right)\right\}\nonumber\\
&&\hphantom{+e^{-3 \lambda} \Bigl[}
-2 e^{\lambda} M^6 q^2 \varphi' \left\{\varphi' \left[2 X
\left(r \lambda'-1\right) f_4'
+ \left(7 r \lambda'-5\right) f_4\right]
-2 r \varphi'' \left(5 f_4+2 X f_4'\right)\right\}\Bigr]\nonumber\\
&&+3 e^{-4 \lambda} M^2 \varphi'^2 \Bigl[e^{\nu}
\left\{\varphi'^2 \left(7 \lambda'
\varphi'-10 \varphi''\right) f_5 -2 e^{\lambda} M^2 X^2 \left(\lambda'
\varphi'-2 \varphi''\right)f_5' \right\}\nonumber\\
&&\hphantom{+3 e^{-4 \lambda} M^2 \varphi'^2 \Bigl[}
-e^{\lambda} M^2 q^2 \left(11 \lambda' \varphi'
-14 \varphi''\right)f_5 \Bigr]\nonumber\\
&&+e^{-3 \lambda} \Bigl[
e^{\lambda} \left\{4 e^{\lambda} M^8 q^2
\left(e^{\lambda}+r \lambda'-1\right)+4
e^{\nu} M^6 \varphi' \left(\varphi'
\left(e^{\lambda}-3 r \lambda'+1\right)+4 r
\varphi''\right)\right\} s_4 \nonumber\\
&&\hphantom{+e^{-3 \lambda} \Bigl[}
+8 M^4 \bigl\{M^4 q^4 e^{2 \lambda -\nu}
\left(e^{\lambda}+r \lambda'-1\right)
-e^{\lambda} \left(e^{\lambda}-1\right) M^2
q^2 \varphi'^2\nonumber\\
&&\hphantom{+e^{-3 \lambda} \Bigl[ +8 M^4 \bigl\{}
+e^{\nu} r \varphi'^3 \left(\lambda'
\varphi'-2 \varphi''\right)\bigr\}s_4' \Bigr]\nonumber\\
&&+e^{-4 \lambda}
\Bigl[
e^{\lambda} \bigl\{6 e^{\nu} M^4 \varphi'^2
\left(\left(e^{\lambda}-5\right) \lambda' \varphi'
-2 \left(e^{\lambda}-3\right)
\varphi''\right)\nonumber\\
&&\hphantom{+e^{-4 \lambda} \Bigl[ e^{\lambda} \bigl\{}
-6 e^{\lambda} M^6 q^2 \left(\left(e^{\lambda}-3\right) \lambda'
\varphi'-2 \left(e^{\lambda}-1\right)
\varphi''\right)\bigr\}s_5 \nonumber\\
&&\hphantom{+e^{-4 \lambda} \Bigl[}
+e^{-\nu}
\bigl\{-4 e^{2 \lambda} M^6 q^4 \left(\left(e^{\lambda}-3\right)
\lambda' \varphi'
-2 \left(e^{\lambda}-1\right) \varphi''\right)\nonumber\\
&&\hphantom{+e^{-4 \lambda} \Bigl[ +e^{-\nu} \bigl\{}
+8 \left(e^{\lambda}-1\right) M^4 q^2
e^{\lambda +\nu} \varphi'^2 \left(\lambda'
\varphi'-2 \varphi''\right)\nonumber\\
&&\hphantom{+e^{-4 \lambda} \Bigl[ +e^{-\nu} \bigl\{}
-4 \left(e^{\lambda}-3\right) e^{2 \nu} M^2
\varphi'^4 \left(\lambda' \varphi'-2 \varphi''\right)\bigr\}s_5'
\Bigr]\nonumber\\
&&= M^6 M_\text{Pl}^2 e^{\nu -\lambda} \left[1-r \lambda'-e^{\lambda}
\left(1-\Lambda_\text{bare} r^2\right)\right],
\label{B1}
\end{eqnarray}
where $X = e^{-\nu} q^2 - e^{-\lambda}\varphi'^2/M^2$, and where
the primes denote derivatives with respect to the argument of the
corresponding terms, i.e., $f_n' = df_n(X)/dX$ and
$s_n' = ds_n(X)/dX$, but
$\varphi' = \partial_r\varphi$, $\varphi'' = \partial_r^2\varphi$,
$\lambda' = \partial_r\lambda$ and $\nu' = \partial_r\nu$.

The second equation expresses that the linear combination
(\ref{eqCombLinEinsteinCurrent}) vanishes for $\mu=\nu = r$, and
we multiply it by a global factor
$e^{2\lambda} M^4 M_\text{Pl}^2 r^2$:
\begin{eqnarray}
&&e^{\lambda} M^8 r^2 X f_2\nonumber\\
&&+e^{-2 \lambda -\nu}
\left[2 e^{\nu} M^2 \varphi'^4 \left(r \nu'+1\right)-2 e^{\lambda}
M^4 q^2 \varphi' \left(\varphi' \left(-r \lambda'
+2 r \nu'+1\right)+2 r \varphi''\right)\right]
f_4\nonumber\\
&&+6 e^{-3 \lambda -\nu} \varphi'^2
\left[e^{\nu} \nu' \varphi'^3+e^{\lambda} M^2 q^2 \left(\varphi'
\left(\lambda'-2 \nu'\right)-2 \varphi''\right)\right]
f_5\nonumber\\
&&+
\left[4 e^{-\nu} M^6 q^2 \left(e^{\lambda}+r \nu'-1\right)
-4 e^{-\lambda} M^4
\varphi'^2 \left(e^{\lambda}+r \nu'+1\right)\right]
s_4\nonumber\\
&&+8 M^4 q^2 r
e^{-\lambda -2 \nu} \left[e^{\lambda} M^2 q^2 \nu'
-e^{\nu} \varphi'
\left(\lambda' \varphi'-2 \varphi''\right)\right]
s_4' \nonumber\\
&&+4 M^2 e^{-2
(\lambda +\nu )} \varphi' \left[e^{\lambda} M^2 \nu'
\left(3 e^{2 \nu} X s_5+2 q^4 s_5'\right)
-2 e^{\nu} q^2 \varphi' \left(\lambda'
\varphi'-2 \varphi''\right)s_5' \right]\nonumber\\
&&= -M^4 M_\text{Pl}^2 \left[1+r \nu'-e^{\lambda}
\left(1-\Lambda_\text{bare} r^2\right)\right].
\label{B2}
\end{eqnarray}
Note that no derivative of any function $f_n$ enters this linear
combination~(\ref{eqCombLinEinsteinCurrent}), although some
$s_4'$ and $s_5'$ do remain, as underlined at the end of
Sec.~\ref{Sec2}. Note in particular that the function $f_3$ fully
disappears from this combination. The reason is that the same
term $\propto (3 f_3+2 X f_3')$ enters both the $rr$-component of
the Einstein equations and the scalar current, and we know that
$f_3'$ must cancel in the
combination~(\ref{eqCombLinEinsteinCurrent}).

The third equation is the radial component of the scalar current
(\ref{eqDefCurrent}), globally multiplied by a factor
$-2 e^{\lambda} M^6 r^2$:
\begin{eqnarray}
&&4 M^8 r^2 \varphi' \left(f_2+X f_2'\right)\nonumber\\
&&+ M^6 r
e^{-\lambda -\nu} \left(3 f_3+2 X f_3'\right) \left[e^{\nu}
\varphi'^2 \left(r \nu'+4\right)
-e^{\lambda} M^2 q^2 r \nu'\right]\nonumber\\
&&+ e^{-2 \lambda -\nu} \left[8 e^{\nu} M^4 \varphi'^3
\left(r \nu'+1\right)\left(2 f_4+X f_4'\right)-4 e^{\lambda
} M^6 q^2 r \varphi' \left(\nu' \left(5 f_4+2 X f_4'\right)
+\lambda' f_4\right)\right]\nonumber\\
&&-6 M^2 e^{-3 \lambda -\nu} \varphi'^2
\left[e^{\lambda} M^2 q^2 \left(2 \lambda'+7\nu'\right) f_5
-e^{\nu} \nu' \left(5 \varphi'^2 f_5
-2 e^{\lambda} M^2 X^2 f_5'\right)\right]\nonumber\\
&&+16 M^4 e^{-2 \lambda -\nu}
\varphi' \left[e^{\lambda} M^2 q^2 \left(e^{\lambda}
+r \lambda'-1\right)s_4' +e^{\nu}
\left(e^{\lambda}-r \nu'-1\right) \left(e^{\lambda} M^2
s_4-\varphi'^2 s_4' \right)\right]\nonumber\\
&&+ e^{-3 \lambda -2 \nu}
\Bigl[
-8 e^{2 \lambda} \left(e^{\lambda}-1\right) M^6 q^4 \nu' s_5'
+4 \left(e^{\lambda}-3\right) e^{2 \nu} M^2 \nu' \varphi'^2
\left(3 e^{\lambda} M^2 s_5-2 \varphi'^2 s_5'\right)\nonumber\\
&&\hphantom{+ e^{-3 \lambda -2 \nu} \Bigl[}
-4 M^4 q^2 e^{\lambda +\nu} \left\{\left(e^{\lambda}-1\right)
\nu' \left(3 e^{\lambda} M^2 s_5-4 \varphi'^2 s_5' \right)-4
\lambda' \varphi'^2 s_5'\right\}\Bigr]= 0.
\label{B3}
\end{eqnarray}


\begin{thebibliography}{99}

\bibitem{Martin:2012bt}
J.~Martin,
Comptes Rendus Physique {\bf 13} (2012) 566
[arXiv:1205.3365 [astro-ph.CO]].

\bibitem{Koksma:2011cq}
J.~F.~Koksma and T.~Prokopec,
arXiv:1105.6296 [gr-qc].

\bibitem{Dolgov:1997za}
A.~D.~Dolgov,
in \textit{Paris 1997, Phase transitions in cosmology}, 161--175
[astro-ph/9708045].

\bibitem{Horndeski}
G.~W.~Horndeski,
Int. J. Theor. Phys. {\bf 10}, 363 (1974).

\bibitem{Nicolis:2008in}
A.~Nicolis, R.~Rattazzi and E.~Trincherini,
Phys. Rev. D {\bf 79}, 064036 (2009)
[arXiv:0811.2197 [hep-th]].

\bibitem{Deffayet:2009wt}
C.~Deffayet, G.~Esposito-Far\`ese and A.~Vikman,
Phys. Rev. D {\bf 79} (2009) 084003
[arXiv:0901.1314 [hep-th]].

\bibitem{Deffayet:2009mn}
C.~Deffayet, S.~Deser and G.~Esposito-Far\`ese,
Phys. Rev. D {\bf 80} (2009) 064015
[arXiv:0906.1967 [gr-qc]].

\bibitem{Deffayet:2011gz}
C.~Deffayet, X.~Gao, D.~A.~Steer and G.~Zahariade,
Phys. Rev. D {\bf 84} (2011) 064039
[arXiv:1103.3260 [hep-th]].

\bibitem{Zumalacarregui:2013pma}
M.~Zumalac\'arregui and J.~Garc\'{\i}a-Bellido,
Phys.\ Rev.\ D {\bf 89}, 064046 (2014)
[arXiv:1308.4685 [gr-qc]].

\bibitem{Gleyzes:2014dya}
J.~Gleyzes, D.~Langlois, F.~Piazza and F.~Vernizzi,
Phys. Rev. Lett. {\bf 114} (2015) 211101
[arXiv:1404.6495 [hep-th]];

\bibitem{Gleyzes:2014qga}
J.~Gleyzes, D.~Langlois, F.~Piazza and F.~Vernizzi,
JCAP {\bf 1502} (2015) 018
[arXiv:1408.1952 [astro-ph.CO]].

\bibitem{Lin:2014jga}
C.~Lin, S.~Mukohyama, R.~Namba and R.~Saitou,
JCAP {\bf 1410} (2014) 071
[arXiv:1408.0670 [hep-th]];

\bibitem{Deffayet:2015qwa}
C.~Deffayet, G.~Esposito-Far\`ese and D.~A.~Steer,
Phys. Rev. D {\bf 92} (2015) 084013
[arXiv:1506.01974 [gr-qc]].

\bibitem{Langlois:2015cwa}
D.~Langlois and K.~Noui,
JCAP {\bf 1602} (2016) 034
[arXiv:1510.06930 [gr-qc]].

\bibitem{Crisostomi:2016czh}
M.~Crisostomi, K.~Koyama and G.~Tasinato,
JCAP {\bf 1604} (2016) 044
[arXiv:1602.03119 [hep-th]].

\bibitem{deRham:2016wji}
C.~de Rham and A.~Matas,
JCAP {\bf 1606} (2016) 041
[arXiv:1604.08638 [hep-th]].

\bibitem{BenAchour:2016fzp}
J.~Ben Achour, M.~Crisostomi, K.~Koyama, D.~Langlois, K.~Noui and G.~Tasinato,
arXiv:1608.08135 [hep-th].

\bibitem{Charmousis:2011bf}
C.~Charmousis, E.~J.~Copeland, A.~Padilla and P.~M.~Saffin,
Phys. Rev. Lett. {\bf 108} (2012) 051101
[arXiv:1106.2000 [hep-th]].

\bibitem{Charmousis:2011ea}
C.~Charmousis, E.~J.~Copeland, A.~Padilla and P.~M.~Saffin,
Phys. Rev. D {\bf 85} (2012) 104040
[arXiv:1112.4866 [hep-th]].

\bibitem{Babichev:2015qma}
E.~Babichev, C.~Charmousis, D.~Langlois and R.~Saito,
Class. Quant. Grav. {\bf 32} (2015) 242001
[arXiv:1507.05942 [gr-qc]].

\bibitem{Appleby:2012rx}
S.~A.~Appleby, A.~De Felice and E.~V.~Linder,
JCAP {\bf 1210} (2012) 060
[arXiv:1208.4163 [astro-ph.CO]].

\bibitem{Linder:2013zoa}
E.~V.~Linder,
JCAP {\bf 1312} (2013) 032
[arXiv:1310.7597 [astro-ph.CO]].

\bibitem{Starobinsky:2016kua}
A.~A.~Starobinsky, S.~V.~Sushkov and M.~S.~Volkov,
JCAP {\bf 1606} (2016) 007
[arXiv:1604.06085 [hep-th]].

\bibitem{Martin-Moruno:2015bda}
P.~Mart\'{\i}n-Moruno, N.~J.~Nunes and F.~S.~N.~Lobo,
Phys. Rev. D {\bf 91} (2015) no.08, 084029
[arXiv:1502.03236 [gr-qc]].

\bibitem{Martin-Moruno:2015lha}
P.~Mart\'{\i}n-Moruno, N.~J.~Nunes and F.~S.~N.~Lobo,
JCAP {\bf 1505} (2015) 033
[arXiv:1502.05878 [gr-qc]].

\bibitem{Babichev:2012re}
E.~Babichev and G.~Esposito-Far\`ese,
Phys. Rev. D {\bf 87} (2013) 044032
[arXiv:1212.1394 [gr-qc]].

\bibitem{Babichev:2013usa}
E.~Babichev and C.~Deffayet,
Class. Quant. Grav. {\bf 30} (2013) 184001
[arXiv:1304.7240 [gr-qc]].

\bibitem{Babichev:2013cya}
E.~Babichev and C.~Charmousis,
JHEP {\bf 1408} (2014) 106
[arXiv:1312.3204 [gr-qc]].

\bibitem{Cisterna:2015yla}
A.~Cisterna, T.~Delsate and M.~Rinaldi,
Phys. Rev. D {\bf 92} (2015) no.4, 044050
[arXiv:1504.05189 [gr-qc]].

\bibitem{Appleby:2015ysa}
S.~Appleby,
JCAP {\bf 1505} (2015) 009
[arXiv:1503.06768 [gr-qc]].

\bibitem{Langlois:2015skt}
D.~Langlois and K.~Noui,
JCAP {\bf 1607} (2016) 016
[arXiv:1512.06820 [gr-qc]].

\bibitem{Crisostomi:2016tcp}
M.~Crisostomi, M.~Hull, K.~Koyama and G.~Tasinato,
JCAP {\bf 1603} (2016) 038
[arXiv:1601.04658 [hep-th]].

\bibitem{Kobayashi:2014eva}
T.~Kobayashi and N.~Tanahashi,
PTEP {\bf 2014}, 073E02 (2014)
[arXiv:1403.4364 [gr-qc]].

\bibitem{Deffayet:2010zh}
C.~Deffayet, S.~Deser and G.~Esposito-Far\`ese,
Phys. Rev. D {\bf 82}, 061501 (2010)
[arXiv:1007.5278 [gr-qc]].

\bibitem{Misner:1974qy}
C.~W.~Misner, K.~S.~Thorne and J.~A.~Wheeler,
\textit{Gravitation}, Freeman, San Francisco (1973).

\bibitem{Kobayashi:2011nu}
T.~Kobayashi, M.~Yamaguchi and J.~Yokoyama,
Prog. Theor. Phys. {\bf 126}, 511 (2011)
[arXiv:1105.5723 [hep-th]].

\bibitem{Ostrogradski} M. Ostrogradski,
Mem. Ac. St. Petersbourg 4, 385 (1850).

\bibitem{DeFelice:2015sya}
A.~De Felice, R.~Kase and S.~Tsujikawa,
Phys. Rev. D {\bf 92} (2015) no.12, 124060
[arXiv:1508.06364 [gr-qc]].

\bibitem{Babichev:2009ee}
E.~Babichev, C.~Deffayet and R.~Ziour,
Int. J. Mod. Phys. D {\bf 18} (2009) 2147
[arXiv:0905.2943 [hep-th]].

\bibitem{Kimura:2011dc}
R.~Kimura, T.~Kobayashi and K.~Yamamoto,
Phys. Rev. D {\bf 85} (2012) 024023
[arXiv:1111.6749 [astro-ph.CO]].

\bibitem{Narikawa:2013pjr}
T.~Narikawa, T.~Kobayashi, D.~Yamauchi and R.~Saito,
Phys. Rev. D {\bf 87} (2013) 124006
[arXiv:1302.2311 [astro-ph.CO]].

\bibitem{DeFelice:2011th}
A.~De Felice, R.~Kase and S.~Tsujikawa,
Phys. Rev. D {\bf 85} (2012) 044059
[arXiv:1111.5090 [gr-qc]].

\bibitem{Kase:2013uja}
R.~Kase and S.~Tsujikawa,
JCAP {\bf 1308} (2013) 054
[arXiv:1306.6401 [gr-qc]].

\bibitem{Koyama:2013paa}
K.~Koyama, G.~Niz and G.~Tasinato,
Phys. Rev. D {\bf 88} (2013) 021502
[arXiv:1305.0279 [hep-th]].

\bibitem{Charmousis:2015aya}
C.~Charmousis and D.~Iosifidis,
J. Phys. Conf. Ser. {\bf 600} (2015) 012003
[arXiv:1501.05167 [gr-qc]].

\bibitem{Babichev:2011iz}
E.~Babichev, C.~Deffayet and G.~Esposito-Far\`ese,
Phys. Rev. Lett. {\bf 107}, 251102 (2011)
[arXiv:1107.1569 [gr-qc]].

\bibitem{Kobayashi:2014ida}
T.~Kobayashi, Y.~Watanabe and D.~Yamauchi,
Phys. Rev. D {\bf 91} (2015) 064013
[arXiv:1411.4130 [gr-qc]].

\bibitem{Babichev:2016jom}
E.~Babichev, K.~Koyama, D.~Langlois, R.~Saito and J.~Sakstein,
arXiv:1606.06627 [gr-qc].

\bibitem{Babichev:2015rva}
E.~Babichev, C.~Charmousis and M.~Hassaine,
JCAP {\bf 1505} (2015) 031
[arXiv:1503.02545 [gr-qc]].

\bibitem{Hui:2012qt}
L.~Hui and A.~Nicolis,
Phys. Rev. Lett. {\bf 110} (2013) 241104
[arXiv:1202.1296 [hep-th]].

\bibitem{Babichev:2016rlq}
E.~Babichev, C.~Charmousis and A.~Leh\'ebel,
Class. Quant. Grav. {\bf 33} (2016) no.15, 154002
[arXiv:1604.06402 [gr-qc]].

\bibitem{Babichev:2016fbg}
E.~Babichev, C.~Charmousis, A.~Leh\'ebel and T.~Moskalets,
arXiv:1605.07438 [gr-qc].

\end{thebibliography}
\end{document}